\documentclass[sn-mathphys]{sn-jnl}

\usepackage[utf8]{inputenc}
\usepackage{amsmath,amssymb}
\usepackage{outlines}
\usepackage{subcaption}

\usepackage{latexsym}
\usepackage{amsfonts}
\usepackage{siunitx}
\usepackage{gensymb}
\usepackage{tabu}
\usepackage{mathtools}
\usepackage{natbib}
\usepackage{graphicx}
\usepackage{caption}
\usepackage{fancyhdr} 
\usepackage{makeidx}
\usepackage{tabularx}
\usepackage{makecell}
\usepackage{xfrac}
\usepackage{xcolor}
\usepackage{makecell}
\usepackage{tabularx}
\usepackage{graphicx}
\usepackage{adjustbox} 

\usepackage{url}
\definecolor{newcolor}{rgb}{.8,.349,.1}

\jyear{2022}%

\theoremstyle{thmstyleone}%
\theoremstyle{thmstyletwo}%

\theoremstyle{thmstylethree}%

\begin{document}

\title[A detailed dynamical model for inclination-only dependent lunisolar resonances.]{A detailed dynamical model for inclination-only dependent lunisolar resonances. Effect on the "eccentricity growth" mechanism.}


\author*[1]{\fnm{Edoardo} \sur{Legnaro}}\email{legnaro@academyofathens.gr}

\author[2]{\fnm{Christos} \sur{Efthymiopoulos}}\email{christos.efthymiopoulos@math.unipd.it}

\affil*[1]{\centering \scriptsize \orgdiv{Department of Physics}, \orgname{Aristotle University of Thessaloniki}, \\ \orgaddress{\city{Thessaloniki}, \postcode{54124}, \country{Greece}}\\
\orgdiv{Research Center for Astronomy and Applied Mathematics}, \orgname{Academy of Athens}, \orgaddress{\city{Athens}, \postcode{11527}, \country{Greece}}}

\affil[2]{\scriptsize \orgdiv{Department of Mathematics}, \orgname{Università degli Studi di Padova}, \\ \orgaddress{\street{Via Trieste 63}, \city{Padova}, \postcode{35121}, \country{Italy}}}


\abstract{ \small The focus of this paper is on inclination-only dependent lunisolar resonances, which shape the dynamics of a MEO (Medium Earth Orbit) object over secular time scales (i.e. several decades). 
Following the formalism of \cite{daquin2021deep}, we discuss an analytical model yielding the correct form of the separatrices of each one of the major lunisolar resonances in the "action" space $(i, e)$ (inclination, eccentricity) for any given semi-major axis $a$. We then highlight how our method is able to predict and explain the main structures found numerically in Fast Lyapunov Indicator (FLI) cartography. We focus on explaining the dependence of the FLI maps from the initial phase of the argument of perigee $\omega$ and of the longitude of the ascending node $\Omega$ of the object and of the moon $\Omega_L$.
In addition, on the basis of our model, we discuss the role played by the $\Omega-\Omega_L$ and the $2 \Omega-\Omega_L$ resonances, which overlap with the inclination-only dependent ones as they sweep the region for increasing values of $a$, generating large domains of chaotic motion.
Our results provide a framework useful in designing low-cost satellite deployment or space debris mitigation strategies, exploiting the natural dynamics of lunisolar resonances that increase an object's eccentricity up until it reaches a domain where friction leads to atmospheric re-entry. }




\maketitle

\clearpage

\section{Introduction}
The region on which we focus our attention is delimited by semi-major axis $ a \in [17000$ $\si{\km},$ $ 32000$ $\si{\km} ]$, eccentricities $e \in [0,1)$ and inclinations $ i \in [30 \degree$, $90 \degree]$. 
This region is of particular interest from both a theoretical and practical standpoint.
Indeed, it is very rich from a dynamical point of view, due to the presence of a web of numerous secular and semi-secular lunisolar resonances ( see \cite{hughes1980earth}), which causes chaotic motion. See \cite{rossi2008resonant}, \cite{rosengren2015chaos}, \cite{daquin2016dynamical}, \cite{celletti2016study}, \cite{gkolias2016order},  and \cite{celletti2017analytical}, where it is possible to find a summary of the most important mathematical models used to study the dynamics of small bodies around the Earth. Also, see \cite{celletti2020resonances} for a guideline to study the geography of the different kinds of resonances.
We recall that the MEO region is home to various artificial satellites, whose most common applications include navigation, communication, and geodetic/space environment science. 
As an example, we have the GNSS (global navigation satellite system) constellations, which have $e\sim 0 $ and consists of the Russian GLONASS $(a \sim 25510 \si{km}, i \sim 64.8 \degree)$, the American GPS  $(a \sim 26561 \si{km}, i \sim 55 \degree)$, the Chinese BEIDOU  $(a \sim 27906 \si{km}, i \sim 55 \degree)$, the European GALILEO  $(a \sim 29601 \si{km}, i \sim 56 \degree)$. In addition, in this region we can find the Molniya satellites, launched by the Soviet Union from 1965 to 2004. These satellites use highly eccentric elliptical orbits (ranging between $e \sim 0.5$ and $e \sim 0.8$ and inclination $ \sim 64 \degree$) known as Molniya orbits, which have a long dwell time over high latitudes. The orbits of all these types are mainly influenced by an inclination-only dependent lunisolar resonance: the $2g+h$ (close to $56^\degree$) or the $2g$ (close to $64^\degree$). 
Recently, many advances in the understanding of Molniya orbits have been made: \cite{alessi2021dynamical}, \cite{daquin2021dynamical}, \cite{talu2021dominant}.

For quite a long time, the problem of lunisolar resonances has remained quite "underrated" (\cite{breiter2001lunisolar}), but it gained more and more interest with the increased awareness of the threat imposed by space debris.   
Indeed, it is possible to design low cost end-of-life (EoL) disposal strategies exploiting resonant effects that increase an object's eccentricity up until it reaches a domain where friction leads to atmospheric re-entry (see \cite{chao2004long}, \cite{rossi2008resonant}, \cite{alessi2014effectiveness}, \cite{alessi2016numerical}, \cite{armellin2018optimal}, \cite{skoulidou2019medium}  and references therein). We will briefly discuss this topic in section \ref{sec:Eccentricity growth}.

Our main contributions in the present work can be summarized as follows.
\begin{itemize}
\item We provide an analytic framework useful to understand the structure of all inclination-only dependent lunisolar resonances and to compute the correct form of their separatrices in the "action" space (i.e. the space of the elements $(e, i)$), as shown in section \ref{sec: FLI Map Prediction}. 
Our analysis extends the one done in \cite{daquin2021deep} by providing a more detailed treatment of the crossing domain of each lunisolar resonance with the $ \Omega - \Omega_L $ and $ 2 \Omega - \Omega_L $ resonances. As an example, we will provide all the details for the particular case of the $2g$ resonance.
Results for all the major lunisolar resonances (the $g+h$, $2g+h$, $2g$, $2g-h$, $g-h$) are summarized in the form of tables with numerical coefficients in the appendix.

\item We show the correspondence between the theoretical phase portraits at each resonance and the numerical stability maps computed by the "Fast Lyapunov Indicator" (FLI, see \cite{froeschle1997fast}, \cite{guzzo2002numerical}).
\item We provide quantitative estimates for the maximum eccentricity, as a function of the initial inclination, which can be reached via the "eccentricity growth" mechanism, separately for each resonance, providing also the limits in inclination within which the mechanism is active, as well as the dependence of these limits on the initial phases $\Omega, \Omega_L$. In fact, analogously to \cite{daquin2021deep}, we identify these limits by the values of the inclination that marks the transition of circular orbits ($e=0$) from stable to unstable via an analogue of the Kozai mechanism (\cite{kozai1962secular}, \cite{lidov1961evolution}).
\end{itemize}

The structure of the paper is as follows: Section \ref{sec: Ham model} discusses the model and basic definitions regarding lunisolar resonances; in Section \ref{sec: Analytic Theory} the details of the construction of our analytic framework are shown; Section \ref{sec: FLI Map Prediction} explains how the correct form of the separatrices of lunisolar resonances can be retrieved from the analytic model; finally, Section \ref{sec:Eccentricity growth} provides estimates for the eccentricity growth mechanism.

\section{The Hamiltonian Model - definition of resonances}
\label{sec: Ham model}
In the present study, we use the same model as in \cite{daquin2021deep}.
Consider a MEO object under the gravitational influence of the Earth, the Moon, and the Sun in an Earth-centered inertial frame. 
We denote as $(a, e, i, M, \omega,  \Omega)$ the orbital elements (semi-major axis, eccentricity, inclination, mean anomaly, argument of the perigee, right ascension of the ascending node).
We assume an Earth-centered inertial frame (ECI coordinates) with the Mean Equator Mean Equinox
convention.
We assume circular orbits for the Moon and the Sun, where the longitude of the solar ascending node is constant: $\Omega_S=0 \; \forall t$.
The system can be modelled by an Hamiltonian of the form
\begin{equation}
    \label{HSEC}
    \mathcal{H}= \mathcal{H}_{k e p}+\mathcal{H}_{J_{2}}+\mathcal{H}_{L S}
\end{equation}
with
\begin{align*}
        \mathcal{H}_{k e p}&=\frac{v^{2}}{2}-\frac{\mu_{E}}{r} \\
        \mathcal{H}_{J_{2}}&=\frac{R_{E}^{2} J_{2} \mu_{E}\left(3 \sin ^{2} \phi-1\right)}{2 r^{3}} \\
        \mathcal{H}_{L S}&=-\frac{\mu_{L}}{r_{L}}\left(\frac{r_{L}}{\left\|\mathbf{r}-\mathbf{r}_{L}\right\|}-\frac{\mathbf{r} \cdot \mathbf{r}_{L}}{r_{L}^{2}}\right)-\frac{\mu_{S}}{r_{S}}\left(\frac{r_{S}}{\left\|\mathbf{r}-\mathbf{r}_{S}\right\|}-\frac{\mathbf{r} \cdot \mathbf{r}_{S}}{r_{S}^{2}}\right),
\end{align*}

where $r,v$ denotes the geocentric distance and velocity of the test-particle, $\phi$ is the geocentric latitude, $R_E$ is the mean equatorial Earth’s radius and $J_2$ the Earth’s oblateness coefficient.
The geocentric vectors of the Moon and Sun are, respectively, $\mathbf{r}_{L}$ and $\mathbf{r}_{S}$.
The gravitational parameters for the Earth, the Moon, and the Sun are respectively $\mu_E$, $\mu_L$, $\mu_S$. Finally, $\left\Vert \bullet \right\Vert
$ denotes the Euclidean norm. 
The numerical values of the physical constants used in this study are provided in Table \ref{tab: num_constants}.

Since we are interested in the secular dynamics, we perform an average on the short-periodic terms, namely those depending on the mean anomalies of the object and the perturbers.
After averaging, the mean anomaly $M$ becomes a cyclic variable, thus its conjugated action $L$ (or equivalently, the semi-major axis $a$) remains constant. 
As a consequence, the \textit{secular Hamiltonian} $\mathcal{H}$ reduces to a time-dependent two degrees of freedom model, in which the dependence on time is just through the longitude of the lunar ascending node  $\Omega_L$, with frequency $\nu_L=-\frac{2\pi}{18.6 \, \text{years}}$.
We introduce a dummy action $A$ conjugated to $\Omega_L$ in order to render the system autonomous.

Thus, the final model reads: 

\begin{equation}
    \bar{\mathcal{H}} = - \frac{\mu_E}{2 a} - \frac{\mu_E J_2 R_E^2}{2 a^3 (1-e^2)^{3/2}} \left( 1- \frac{3}{2} \sin^2 i \right) + \bar{\mathcal{H}}_{LS} + \nu_L A
\end{equation}
The coefficients of the doubly-averaged lunisolar perturbations $\bar{\mathcal{H}}_{LS}$ are given in \cite{kaula1962development} (see also \cite{celletti2017analytical}).
The validation of the model above as a sufficient basis for understanding the dynamics at MEO has been done in several works (\cite{daquin2016dynamical}, \cite{celletti2016bifurcation}, \cite{celletti2017analytical}, \cite{gkolias2019chaotic}).

A \textit{lunisolar secular resonance}, occurs when 
\begin{equation}
\label{resonance}
\exists \left(k_{1}, k_{2}, k_{3}\right) \in \mathbb{Z}^{3} \backslash\{0\} \qquad \text{s. t.} \qquad k_{1} \dot{\omega}+k_{2} \dot{\Omega}+k_{3} \dot{\Omega}_{M}=0,
\end{equation}
where, expressing $\bar{\mathcal{H}}_{J_2}$ in \textit{Delaunay variables} 
$$L=\sqrt{\mu_{E} a}, \; G=L \sqrt{1-e^{2}}, \; H=G \cos i, \; g=\omega, \; h=\Omega,$$ we find (under the $J_2$ approximation)
\begin{equation*}
\dot{\omega} =-\frac{3 R_{E}^{2} J_{2} \mu_{E}^{4}}{4} \frac{1}{L^{3} G^{4}}\left(1-5 \frac{H^{2}}{G^{2}}\right), \quad
\dot{\Omega}=-\frac{3 R_{E}^{2} J_{2} \mu_{E}^{4}}{2} \frac{H}{L^{3} G^{5}},
\end{equation*}
or
\begin{align}
\begin{split}
\dot{\omega} &=\frac{3J_2 R_E^2 \sqrt{a \mu_E} (5 \cos (2 i)+3)}{8 a^4\left(e^2-1\right)^2}, \\
\dot{\Omega} &=-\frac{3 J_2 R_E^2 \sqrt{a \mu_E} \cos i}{2 a^4 \left(e^2-1\right)^2}.
\end{split}
\label{eq: freq_om_OM}
\end{align}
Inserting \eqref{eq: freq_om_OM} into \eqref{resonance} yields an expression involving  the elements $a, e$ and $i$, which provides the location of the secular resonance.
Including terms from the lunisolar tides introduces only a negligible error to this relation at MEO altitude.
From \eqref{eq: freq_om_OM}, we see that resonances of the type $k_1 \dot{\omega}+k_2 \dot{\Omega}$ are independent from  $a$ and $e$. Thus these resonances are called \textit{inclination-only dependent}.
The critical inclination $i_\star$ of the lowest order (and most important) ones is provided in Table \ref{main_res}.

\begin{table}
    \centering
    \caption{Location of the most relevant inclination-only dependent lunisolar resonances.}
    {\renewcommand{\arraystretch}{1.8}
    \begin{tabular}{|c |c |c|}
    \Xhline{1.15pt}  
    Resonance & Critical Inclination $i_\star$ & Location in degrees \\
    \Xhline{1.15pt}  
    $g+h$&  $\arccos \left( \frac{1 + \sqrt{6}}{5}\right)$ &$\sim$ $46.38 \degree$\\
    
    $2g+h$& $\arccos \left( \frac{1 + \sqrt{21}}{10}\right)$ &$\sim$ $56.06 \degree$\\
    
    $2g$& $\frac{1}{2} \arccos \left(- \frac{3}{5}\right)$ &$\sim$  $63.43 \degree$\\
    
    $2g-h$&  $\arccos \left( \frac{-1 + \sqrt{21}}{10}\right)$ &$\sim$ $69.01 \degree$\\
    
    $g-h$&  $\arccos \left( \frac{-1 + \sqrt{6}}{5}\right)$ &$\sim$  $73.15 \degree$\\
    
    $h$& $\frac{\pi}{2}$ &$=$  $90 \degree$ \\
    \Xhline{1.15pt}  
    \end{tabular}}
    \label{main_res}
\end{table}

When dealing theoretically with the problem of lunisolar resonances, an important remark to keep in mind is that all
resonances involving only the frequencies $\nu_L = \dot{\Omega}_L$  and $\dot{\Omega}$ belong to the so-called "first fundamental model" of resonance, thus having the structure of a pendulum, while inclination-only dependent resonances belong, instead, to the "second fundamental model" (see \cite{henrard1983second}, \cite{breiter2001lunisolar}).
This remark is important because the use of the "first fundamental model" (or pendulum) to model the inclination-only dependent resonances leads to very imprecise predictions about the form of each resonance's separatrices (see for example \cite{lei2021secular}).

\begin{figure}[t]
    \centering
    \includegraphics[width= \textwidth]{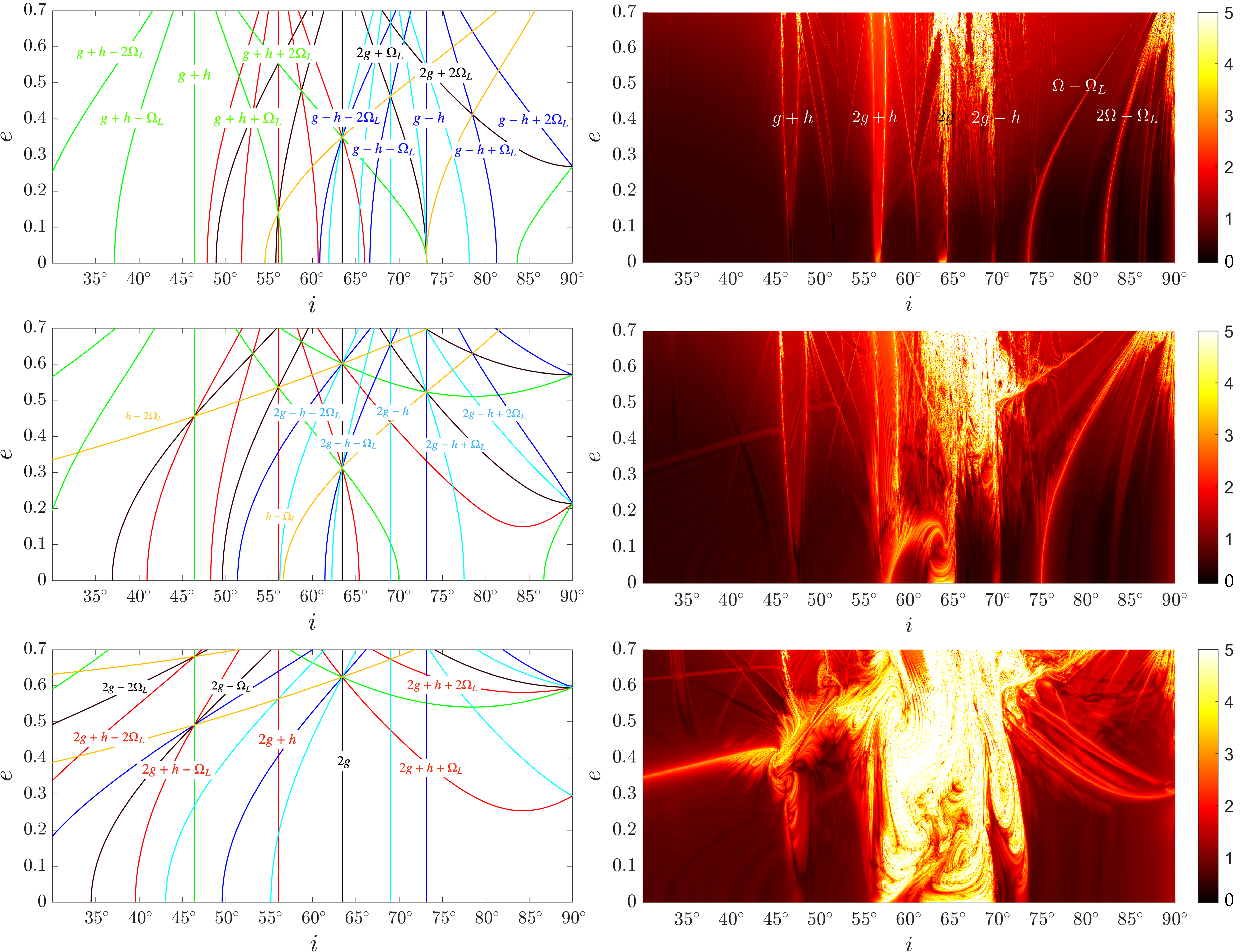}
    \caption{Left: theoretical resonance web obtained by the resonance curves $(i, e)$ for semi-major axis fixed (Eq. \eqref{resonance}-\eqref{eq: freq_om_OM} ) up to order $ \lvert k_1 \rvert + \lvert k_2 \rvert+ \lvert k_3 \rvert = 4$.
    Right:  FLI stability maps for $a = 20000$ $\si{km}$, $24000$ $\si{km}$ and $30000$ $\si{km}$. The remaining initial conditions in the computation of the FLI maps are $\Omega = \omega = \Omega_L = 0$. From the theoretically infinite number of lunisolar resonances just a few are actually important for the dynamics and visible in the FLI maps.
    Notice that the resonances  $\Omega-\Omega_L$ and $2 \Omega-\Omega_L$, while sweeping the phase space as the altitude increases,  overlap with the inclination-only dependent lunisolar resonances and generate larger and larger domains of chaotic motion.}
    \label{fig:res_web}
\end{figure}

To get a global view of how the resonances shape the action space,
Figure \ref{fig:res_web} compares the theoretical web of lunisolar resonances (left) with the FLI map of the same region (right).
The computation of the Fast Lyapunov Indicator is done as in \cite{froeschle1997fast} and \cite{guzzo2002numerical}.
The position of the inclination-dependent resonances listed in Table \ref{main_res}, as well as some of the first order lunar resonances, is marked in the same figure. 
We note that from the theoretically infinite number of lunisolar resonances just a few are actually important for the dynamics, and so visible in the FLI maps.
Besides the inclination-dependent ones, a fundamental role is played by the resonances  $\Omega-\Omega_L$ and $2 \Omega-\Omega_L$.
Indeed, while sweeping the phase space as the altitude increases, they overlap with the inclination-only dependent lunisolar resonances and generate larger and larger domains of chaotic motion (see Figure \ref{fig:res_web}).
As pointed out in \cite{daquin2021deep}, these last two resonances are represented in the Hamiltonian by terms depending on $\sin i_L$, while higher order resonances have coefficients proportional to $\sin^2 i_L$, and so their influence is much smaller. This explain the numerical results found for example in \cite{gkolias2016order} and \cite{rosengren2016galileo}; see also Section \ref{sec: Analytic Theory} below.

\section{Analytic Theory}
\label{sec: Analytic Theory}

\begin{table}[b]    
\centering
\caption{The canonical transformation $f$ for inclination-only dependent lunisolar resonances.}
\begin{adjustbox}{width=\textwidth}
    {\renewcommand{\arraystretch}{1.8}
\begin{tabular}{|c|c|c|c|}
\Xhline{1.15pt}  
Resonance & Gener. Function & Angle variables & Action variables \\
\Xhline{1.15pt}  

$g+h$  & $S=pJ_{R}+q J_{F}$ & $u_R=p, u_F=q$ & $\delta Q=J_F, P=J_R$\\ 
$2g+h$  & $S=\frac{2 p-q}{2} J_{R}+q J_{F}$ & $u_R=p-\frac{q}{2}, u_F=q$ & $\delta Q=J_F-  \frac{J_{R}}{2}, P=J_R$\\

$2g$  & $S=(p-q) J_{R}+q J_{F}$ & $u_R=p-q, u_F=q$ & $\delta Q=J_F- J_R, P=J_R$\\

$2g-h$  & $S=\left( p-\frac{3}{2} q \right) J_{R}+q J_{F}$ & $u_R=p-\frac{3}{2}q, u_F=q$ & $\delta Q=J_{F}-\frac{3}{2} J_{R}, P=J_R$\\

$g-h$  & $S=\left( p -2q \right)J_{R}+q J_{F}$ & $u_R=p-2q, u_F=q$  & $\delta Q=J_{F}-  2 J_{R}, P=J_R$\\
\Xhline{1.15pt}  
\end{tabular}}
\end{adjustbox}
\label{tab:can_transf_f}
\end{table}

In this section, we will discuss in detail the process leading to a useful integrable model for all the lunisolar resonances of Table \ref{main_res}.

\subsection{Resonant Variables}
Consider an inclination-only dependent lunisolar resonance $\sigma=k_1 \omega+k_2\Omega$ (with $k_1 \neq 0$) located at $i_\star$.
First, we need a suitable set of coordinates to study the resonance at hand.
We start with \textit{Modified Delaunay variables}
\begin{equation}
P=L(1-\sqrt{1-e^2}), \; Q=G (1-\cos i), \\
p=-\omega -\Omega, \; q=-\Omega.
\end{equation}

First, we perform a Taylor expansion up to a preselected order $n$ around $\left(e=0,i=i_\star \right)$. 
This is done by performing in $\mathcal{H}$ the substitutions $Q \rightarrow Q_\star + \varepsilon \, \delta Q, \; P \rightarrow  \varepsilon \, \delta P,$
where $\varepsilon$ is a \textit{bookkeeping}  parameter. After the expansion, we set $\varepsilon=1$, and recover a polynomial truncated series form for the Hamiltonian in the variables $\delta Q, \delta P$ .
Next we introduce a suitable canonical transformation 
$$ (\delta P, \delta Q, p, q) \overset{f}{\longmapsto}  (J_R, J_F, u_R, u_F)$$ 
to resonant variables, where $u_R$ is related to the resonant angle $k_1 \omega + k_2 \Omega$ and $u_F$ is a fast angle. 
We build $f$ as the canonical transformation generated by the function $S=-\sigma/2 J_R+q J_F$. Here, depending on the selected resonance for study, $\sigma$ could be $2g+2h$, $2g-h$, $2g$, $2g-h$ or $2g-2h$.
Table \ref{tab:can_transf_f} summarizes the explicit expression of $f$ for the lowest order inclination dependent resonances.

Finally, by introducing \textit{Poincaré variables} $X=\sqrt{2 J_R} \sin u_R$, $Y=\sqrt{2 J_R} \cos u_R,$
 we arrive at a resonant form of the Hamiltonian $H (X,Y,J_F,u_F,A,\Omega_L)$.
The canonical change of variables $f$ is chosen so that $H$ is a polynomial in the variables $X$ and $Y$.

Now, performing the above steps, we readily see that the Hamiltonian $H$ has the structure $H=H_R+H_{CM}+H_M$
with $H_{CM}=H_{CM}^0+H_{CM}^1$, where:
\begin{align}
\begin{split}
    &H_R=R_{20} X^{2}+R_{02} Y^{2}+R_{22} X^{2} Y^{2}+R_{40} X^{4}+R_{04} Y^{4}+\ldots \\
&H_{C M}^{0} =\omega_F J_{F}+\alpha J_{F}^{2}+C_1 \cos u_{F}+C_2 \cos 2 u_{F}+\ldots \\
&H_{C M}^{1} = A \, \nu_L+D_{21} \cos \left(2 u_{F}+\Omega_L \right)+D_{11} \cos \left(u_{F}+\Omega_L \right)+\ldots \\
&H_M = M_{120} J_{F} X^{2}+M_{102} J_{F} Y^{2}+\ldots
\end{split}
\end{align}

\begin{figure}
    \centering
    \includegraphics[width=\textwidth]{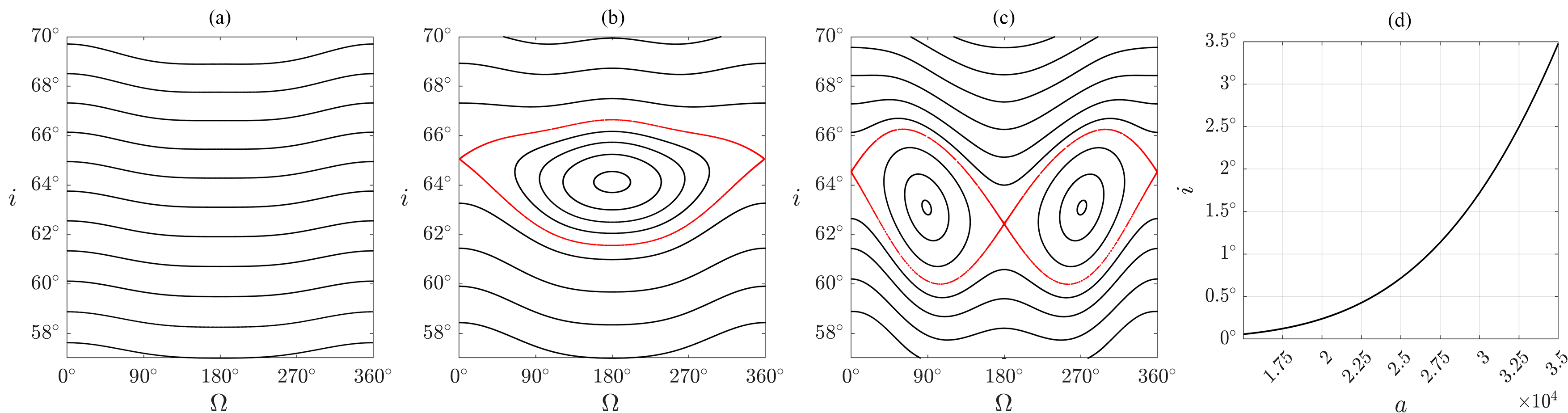}
    \caption{From left to right: Center Manifold for the $2g$ resonance at $20000$ $\si{km}$, $22500$  $\si{km}$, $28000$ $\si{km}$ and inclination of the Laplace Plane between $15000$ $\si{km}$ and $35000$ $\si{km}$. We can see the influence of the resonances $\Omega-\Omega_L$ (b) and $2 \Omega-\Omega_L$ (c) on the Center Manifold.
    In order to numerically visualize it, we compute the Poincaré map of $H_{CM}$ for a surface where $\Omega_L=0$.}
    \label{Fig:CM_Laplace_Plane}
\end{figure}

The numerical values of the constants appearing in the Hamiltonian $R_{20}, R_{02}, R_{22}, R_{40}, R_{04}, C_1, C_2, D_{21}, D_{11}, \alpha$ and $\omega_F$ can be found in the appendix.
The indices "$R$" and "$CM$" in the above notation stand for "Resonant" and "Center Manifold" respectively. The precise meaning of these terms is discussed in \cite{daquin2021deep}, and also in detail below. On the other hand, the terms in $H_M$ instead represent a coupling between resonant $(X, Y)$ and "Center Manifold" variables $(u_F, J_F, \Omega_L, A)$, which play a key role in the appearance of chaos in the system.
Regarding in particular the terms $H_R$ and $H_{CM}$, we have that:
\begin{itemize}
    \item the term $H_R$ yields a 1DOF integrable Hamiltonian related to the center of the particular resonance at hand,
    \item the term $H_{CM}$ describes the dynamics of circular orbits.
Indeed, if $e=0$ it holds $X=0$ and $Y=0$. But for these values of $X$ and $Y$ we have $\dot{X}=H_Y=0$ and $\dot{Y}=-H_X=0$ $\forall t$, since we don't have linear terms in $X$ and $Y$.
It follows that the dynamics of circular orbits is described by $H_{CM}$, and the motion takes place in an invariant sub-manifold embedded in the phase space, the \textit{center manifold}.
In particular, $H_{CM}^0$ is an integrable Hamiltonian describing the evolution of the inclination vector around a fixed point (forced inclination) corresponding to the inclination of the Laplace plane, usually defined as the
plane normal to the axis about which the pole of a satellite’s orbit precesses (\cite{allan1964long}, \cite{kudielka1997equilibria}, \cite{tremaine2009satellite}). Thus the phase space under the flow of $H_{CM}^0$ is foliated by rotational tori, the motion along which represents an oscillation around this forced inclination, which increases with the altitude (see Section \ref{sec: the linear approac}, Figure \ref{Fig:CM_Laplace_Plane} and Figure \ref{fig:inc_vect} (a),(d)).
Far from the lunar resonances $u_F+\Omega_L$ and $2u_F+\Omega_L$, the Hamiltonian $H_{CM}^1$ is just a small perturbation of the former dynamics. However, as we approach one of the lunar resonances the phase space changes drastically, as can be seen from Figure \ref{Fig:CM_Laplace_Plane} (b-c).
\end{itemize}

\subsection{An integrable model of the resonance}

As in \cite{daquin2021deep}, we now seek to construct a heuristic integrable model for the resonance, allowing us to derive theoretical forms (and formulas) for the separatrices of each resonance. To this end, however, we now change strategy with respect to \cite{daquin2021deep} and choose a different way to treat the coupling of the center manifold variables $(u_F, J_F)$ with the resonant variables $(X, Y)$ of the problem. In particular, starting with analytic expressions valid for circular orbits, 
our aim is to find an analytic approximation for $J_F(t)$, and then to plug it inside the full Hamiltonian. By averaging over the angles we will end up with an integrable model of the dynamics in the variables $X$ and $Y$.
We now discuss three different approaches to build an analytic approximation of the solution of $J_F(t; a, i_0, \Omega_0, \Omega_L^0)$.

\subsubsection{The Linear Approach}
\label{sec: the linear approac}
Consider a region of the phase space where the resonance in consideration is far from other resonances. 
Setting $X=Y=0$ (which is an invariant sub-manifold of the phase space under the full Hamiltonian flow), the Hamiltonian governing the dynamics of circular orbits is just $H_{CM}^0$. As in \cite{daquin2021deep}, we call the sub-manifold $X=0=Y$ the \textit{center manifold}.
Far from other resonances, the center manifold is foliated by rotational tori, describing small oscillations with amplitude equal to twice the inclination of the Laplace plane (see \cite{daquin2021deep} and Figure \ref{Fig:CM_Laplace_Plane} (a)).
An accurate model for $J_F(t)$ can then be constructed by considering only the main  terms in $H_{CM}^0$. However, close to the crossing of the considered resonance with the $\Omega-\Omega_L$ or the $2\Omega-\Omega_L$ resonances, the Hamiltonian describing the center manifold dynamics has to be modified with respect to $H_{CM}^0$ according to
\begin{align}
\begin{split}
    K=& \; \omega_F J_{F}+\alpha J_{F}^{2}+C_1 \cos u_{F}+C_2 \cos 2 u_{F}+ \\
 &\quad + D_{11} \cos ( u_{F}+\Omega_L)+D_{21} \cos  ( 2u_{F}+\Omega_L) 
\end{split}
\end{align}  

\begin{table}[b]
\centering
\caption{The canonical transformation $f$ for inclination-only dependent lunisolar resonances.}
\label{tab: change_of_variables}
{\renewcommand{\arraystretch}{1.8}
\begin{tabular}{|c|c|}
\Xhline{1.15pt}  
$\Omega-\Omega_L$ & $2 \Omega-\Omega_L$ \\
\Xhline{1.15pt}  

$S=(u_F+\Omega_L) J_S+\Omega_L J_L$ &$S=(2 u_F+\Omega_L) J_S+\Omega_L J_L$\\ 

$u_S=u_F+\Omega_L$ & $u_S=2 u_F+\Omega_L$\\

 $J_{FF}=J_S$ & $J_{FF}=2 J_S$\\

 $A=J_S+J_L$ & $A=J_S+J_L$\\
\Xhline{1.15pt}  
\end{tabular}}
\end{table}

We call below \textit{linear approach L} the one in which we neglect altogether the lunar terms $ D_{11} \cos ( u_{F}+\Omega_L), D_{21} \cos  ( 2u_{F}+\Omega_L)$ in $K$. Instead, we will call \textit{linear approach with linear node terms LL} the one in which the terms $ D_{11} \cos ( u_{F}+\Omega_L), D_{21} \cos  ( 2u_{F}+\Omega_L)$ are taken into account.
Far from other resonances $\omega_F$ is more than an order of magnitude bigger with respect to the other coefficients, so we can take
\begin{equation}
    \dot{u}_{F}=\frac{\partial K}{\partial J_{F}} \sim \omega_{F} \; \implies \; u_{F}(t) \sim  u_{F}^0+\omega_{F} t.
\end{equation}

while (setting $\Omega_L(0) = 0)$), $\Omega_L(t)=\nu_L t$ with $\nu_L=\frac{2 \pi}{18.613 \; \text{years}}$.
Substituting the above expressions in the equations of motion for $J_F$ under the Hamiltonian $K$ we find:
\begin{align}
\begin{split}
\dot{J}_{F}=-\frac{\partial K}{\partial u_{F}} & \sim \quad C_{1} \sin u_{F}+2 C_{2} \sin 2 u_{F}+ \\ & +D_{11} \sin ( u_{F}+\Omega_L)+D_{22} \sin  ( 2u_{F}+\Omega_L)  \quad 
\end{split}
\end{align}
So we get
\begin{align}
\begin{split}
J_F(t) &= J_{F}^0+\int_{0}^{t} C_{1}\sin \left(u_{F}(s)\right) d s + \\
 & \qquad + \int_{0}^{t} 2 C_{2} \sin \left(2 u_{F}(s)\right) d s +\dots= \\
&= J_{F}^0+\frac{C_{1}}{\omega_{F}} \cos \left(u_{F}^{0}\right)+\frac{C_{2}}{\omega_{F}} \cos \left(2 u_{F}^{0}\right)+ \\
& \qquad \; \,+\frac{D_{11}}{\omega_{F}+\nu_L} \cos \left(u_{F}^{0}\right)+\frac{2 D_{21}}{2\omega_{F}+\nu_L} \cos \left(2 u_{F}^{0}\right) +\\
& \qquad \; \,-\frac{C_{1}}{\omega_{F}} \cos \left(u_{F}(t)\right)-\frac{C_{2}}{\omega_{F}} \cos \left(2 u_{F}(t)\right)+\\
& \qquad \; \,-\frac{D_{11}}{\omega_{F}+\nu_L} \cos \left(u_{F}(t)+\Omega_L(t)\right)+ \\
& \qquad \; \,-\frac{2 D_{21}}{2 \omega_{F}+\nu_L} \cos \left(2 u_{F}(t)+\Omega_L(t)\right).
\end{split}
\label{eq: JF_an_LL}
\end{align}

\begin{figure}[t]
    \centering
    \includegraphics[width=0.4\textwidth]{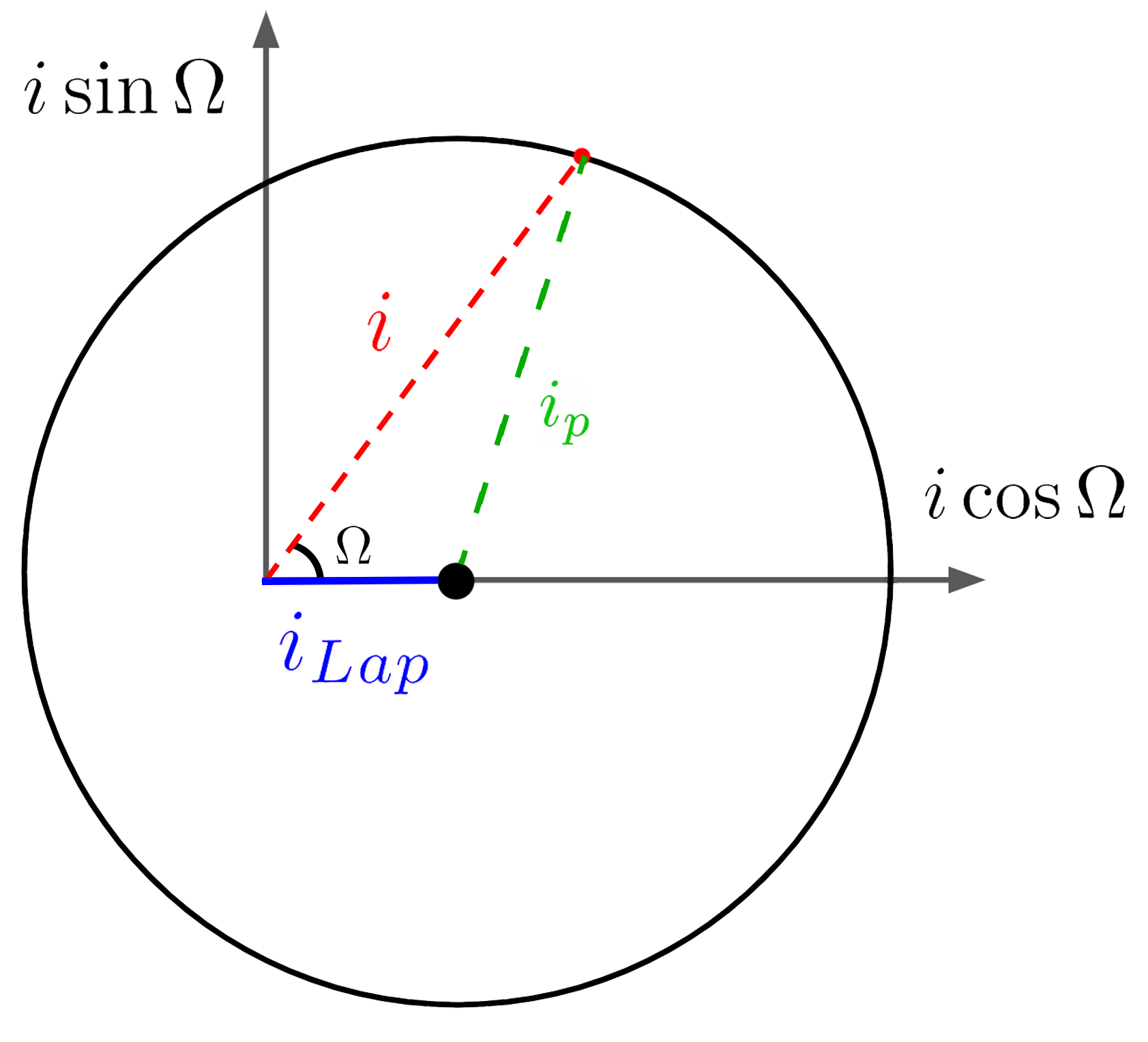}
    \caption{According to the leading terms of \eqref{eq: JF_an_LL}, the inclination vector $(i \cos \Omega, i \sin \Omega)$ follows approximately a circle of radius equal to the proper inclination $i_p$ centered around a point with non-zero inclination equal to the inclination of the Laplace plane $i_{Lap}$.}
    \label{fig:inc_vect}
\end{figure}

The meaning of \eqref{eq: JF_an_LL} can be easily understood with the help of the schematic Figure \ref{fig:inc_vect}. We note that the coefficient $C_1$ is of order $\mathcal{O}(i_E)$, while the coefficient $C_2$ is  $\mathcal{O}(i_E^2)$ (see Table \ref{tab: CM} in the appendix).
The coefficients $D_{11}$ and $D_{21}$ are of order $\mathcal{O}(i_L)$. Thus, the leading terms in \eqref{eq: JF_an_LL} are
\begin{equation}
    J_F(t) = J_F^0+ \frac{C_1}{\omega_F} (\cos( u_F^0)-\cos( u_F(t) )) + \ldots
\end{equation}
Translating this result to orbital elements, we find that at first order (see equation (66) of \cite{daquin2021deep})
\begin{equation}
    i \approx i_p + i_{Lap} \cos \Omega,
\end{equation}
where $i_p$ is the satellite's proper inclination and $i_{Lap}$ is the inclination of the Laplace plane.
So, far from the resonances $\Omega - \Omega_L$ and $2 \Omega - \Omega_L$, the center manifold at first order portraits the motion of the inclination vector whose endpoint describes a closed curve which, up to terms $\mathcal{O}(i_{Lap}^2)$, is a circle with radius $i_p$ and center shifted by a value $i_{Lap}$ (see Figure \ref{Fig:CM_Laplace_Plane}).

Consider now a non-circular orbit. The dynamics is described by the full Hamiltonian
$H (X,Y,J_F,u_F,A,\Omega_L)$.
A useful model is then constructed by plugging into $H$ the solution found above for $J_F$ and $u_F$.
This step is justified as long as $J_F(t), u_F(t)$ remain robust even for non-circular orbits, a fact checked by numerical experiments. Now, for a resonance $\sigma$  the angles $u_F$ and $\Omega_L$ are fast with respect to the resonant angle. So we consider the average of $H$ over this last two angles:
\begin{align}
\begin{split}
\bar{H} &=\frac{1}{2 \pi} \int_{0}^{2 \pi}\left(\frac{1}{2 \pi} \int_{0}^{2 \pi} H d u_{F}\right) d \Omega_{L}= \\
&= \;\gamma_{20} X^{2}+\gamma_{02} Y^{2}+\gamma_{22} X^{2} Y^{2}+\gamma_{40} X^{4}+\gamma_{04} Y^{4}+ \ldots
\end{split}
\end{align}
where the coefficients $\gamma_{ij}$ are functions of $u_F^0,J_F^0, a$.
The Hamiltonian $\bar{H}\left(X, Y ; J_{F}^{0}, u_{F}^{0}, a\right)$ is integrable, and it contains only the resonant variables $X, Y$.

\subsubsection{The Pendulum Approach}
Suppose now to be in a region subjected to the influence of one of the lunar resonances $\Omega-\Omega_L$ or $2 \Omega-\Omega_L$. 
Then, we no longer have a linear evolution of $u_F$ ( Figure \ref{Fig:CM_Laplace_Plane} (b-c)).
In this case, however, we can still construct an analytic model following the method proposed in \cite{gkolias2019chaotic}.
More precisely, we start from a Hamiltonian including only the lunar resonance considered ( $u= u_F + \Omega_L$ or $u= 2 u_F + \Omega_L$)  
$$K=\nu_L A + \omega_F J_{F}+\alpha J_{F}^{2}+C_1 \cos u_{F}+C_2 \cos 2 u_{F}+ D \cos u $$
where $ D \cos u$ stands for $D_{11} \cos ( u_{F}+\Omega_L)$ or $D_{21} \cos  ( 2u_{F}+\Omega_L)$.)

Then we apply the canonical transformation
$$J_F\mapsto J_{FF} - \frac{C_1}{\omega_F} \cos u_F -  \frac{C_2}{2 \omega_F} \cos 2u_F.$$
The new harmonic terms generated from $J_F^2$ have a minimal contribution to the dynamics, thus we can neglect them.
We end up with the Hamiltonian
$K'=\omega_F J_{FF} + \alpha J_{FF}^2+D \cos u.$
We then perform another canonical transformation $(u_F,J_{FF}, A) \mapsto (u_S, J_S,J_L)$ with generating function $S=u J_S + \Omega_L J_L$ (see Table \ref{tab: change_of_variables}).

In the resulting Hamiltonian, $J_L$ is a constant of motion which can be set equal to $0$ without loss of generality.
We end up with the Hamiltonian of a pendulum with the equilibrium point $J_S^\star$ translated from the origin. 
So, with a last transformation $J_S \mapsto \delta J_S+J_S^\star$ we arrive at a Hamiltonian $H_P$ of a pendulum with origin in $0$ in the variables $\delta J_S, u_S$.
We can then express  $\delta J_S$ has a function of the energy $E$, choosing the right sign of the square root in the solution with respect to being in a region above or below the separatrix of the lunar resonance at hand.
We can now go back following the transformations $\delta J_S \mapsto J_S \mapsto J_{FF} \mapsto J_F$, finding an expression $J_F(u_F,E)$. Finally, by imposing initial conditions $J_F^0=J_F(u_F^0,E_0)$,  solving for $E_0$ and then substituting the value found we can reach an expression for $J_F$ as a function of the initial conditions.

At this point we can plug in the Hamiltonian the solution for $J_F$ found and proceed with the computation of the average over $u_F$ and $\Omega_L$, arriving at an integrable model of the dynamics even close to one of the lunar resonances considered.
We will call this method the \textit{pendulum approach} \textit{P}.
For more details see \cite{gkolias2019chaotic}.

\subsubsection{Comparison of the Methods}
\begin{figure}
    \centering
    \includegraphics[width = 0.7\textwidth]{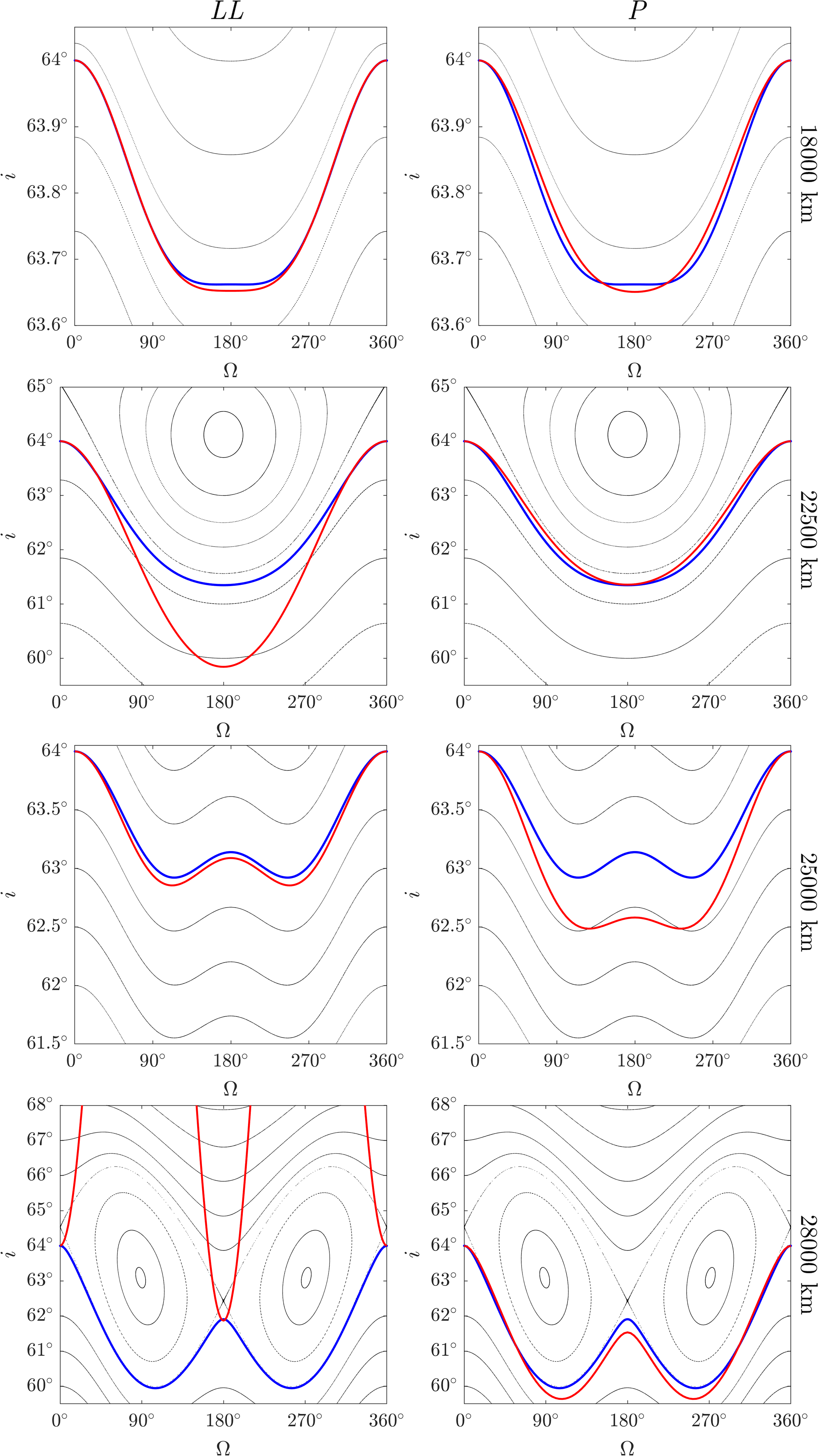}
    \caption{$2g$ resonance: numerical (blue) and analytical (red) solution of $J_F$ at $18000$ $\si{km}$, $22500$ $\si{km}$, $25000$ $\si{km}$ and $28000$ $\si{km}$ comparing the \textit{LL} approach (left) with the \textit{P} one (right).
    Below $16000$ km, the \textit{L} approach would work best. At $18000$ $\si{km}$ the influence of the $\Omega-\Omega_L$ is still small, so the \textit{LL} and the \textit{P} approach are comparable.
    Around $22500$ $\si{km}$ because of the $\Omega-\Omega_L$ resonance the \textit{LL} approach fails while the \textit{P} one works.
    Between $25000$ $\si{km}$ and  $27000$ $\si{km}$, in a domain in-between both resonances, the \textit{P} approach performs worst than the \textit{LL} one.
    Around $28000$ km  we are in the domain of the $2 \Omega-\Omega_L$ resonance, and the \textit{LL} approach fails.}
    \label{fig: tori_2g_LL_vs_P}
\end{figure}

Generally, the pendulum method \textit{P} works best when we are close to the separatrix of the lunar resonance considered, the linear method \textit{L} works better far from it and the $LL$ method works better between the two resonances. 
As an example, for the $2g+h$ resonance the "pendulum" model \textit{P} works better between $23000 \si{km}$ and $25000 \si{km}$  for the $\Omega-\Omega_L$ resonance and between $28000 \si{km}$ and $32000 \si{km}$  for the $2 \Omega-\Omega_L$ one.
This is shown in Figure \ref{fig: tori_2g_LL_vs_P}, where the analytical solution (in red) for $J_F(t)$ is compared against the numerical solution (in blue) obtained without simplifications. We can clearly see that the $P$ method works very well when we are in the domain of the $\Omega-\Omega_L$ or the $2\Omega-\Omega_L$ resonance (second and last row of Figure \ref{fig: tori_2g_LL_vs_P} respectively).
In between, when both resonances influence the dynamics, the method that works best is the $LL$ one, as it considers both such resonances (third row of Figure \ref{fig: tori_2g_LL_vs_P}).
For low altitudes, when the $\Omega-\Omega_L$ and the $2\Omega-\Omega_L$ resonance are quite far, the $L$ approach works better.

\subsection{Bifurcations}
\label{sec: bifurcations}
Following the process described in the previous sections, we derived a 1DOF model depending on the variables $(X, Y)$ and on the parameters $a, J_F^0, u_F^0$, or equivalently $a, i_0, \Omega_0$. We will now discuss the various bifurcations of stable and unstable fixed points altering the form of the phase portraits in $(X, Y)$ as the parameter $J_F^0$ (or $i_0$) changes.

First, fixing a value of $u_F^0$, we are interested in determining the linear stability of the point $X=0$, $Y=0$ (which corresponds to circular orbits).
This amounts to considering the Hessian matrix 
\begin{equation}
\nabla^2 \bar{H} \big \rvert_{X=0, Y=0}  = 
\begin{pmatrix}
\bar{H}_{XX} &  \bar{H}_{XY} \\
\bar{H}_{YX} &  \bar{H}_{YY}
\end{pmatrix} \bigg\rvert_{X=0, Y=0}
\end{equation}
and computing its eigenvalues $\lambda_1$ and $\lambda_2$, which are functions of the parameters $J_F^0$, $a$ and $u_F^0$.
In our case, given the lack of linear terms in $X$ and $Y$, the mixed derivatives will vanish, and so we have $\lambda_1=\bar{H}_{XX }=\gamma_{20}(u_F^0,J_F^0) =\gamma_{20} (i^0, \Omega^0)$ and $\lambda_2=\bar{H}_{YY }=\gamma_{02}(u_F^0,J_F^0) =\gamma_{02} (i^0, \Omega^0)$.

\begin{figure}
    \centering
    \includegraphics[width=\textwidth]{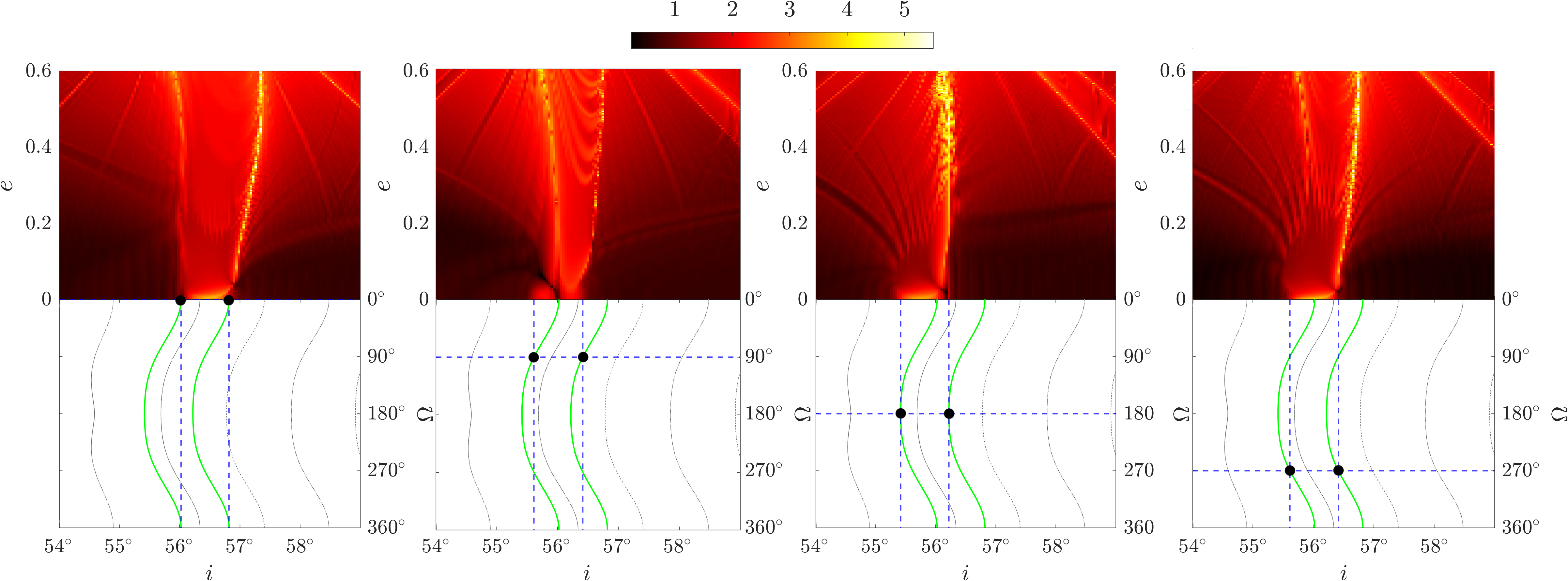}
    \caption{Analytic bounds of the hyperbolic region against an FLI map for the resonance $2g+h$ at $20000$ km. 
    The bounds are shifted along the torus for different values of $\Omega_0$.
    The values of $i_1$ and $i_2$ correspond to the black dots, and the blue dotted line show these values in the FLI map.
    The green lines correspond to the analytic solutions of $J_F(t; a, i_1, \Omega)$ and $J_F(t; a, i_2, \Omega)$.}
    \label{fig:inst_20000_2g_h}
\end{figure}

For $e=0$, $J_F=\delta Q$. Thus $J_F$ becomes a function of the inclination $i$ only, which in turn allows expressing the instability interval as a function of the inclination $i$.
By studying the sign of the function $\lambda_1 \lambda_2$, we can find two bounds within which the point $X=0,$ $Y=0$ is hyperbolic. We will denote such limiting values $i_1$ and $i_2$.
Since, as already remarked, a circular orbit belongs to one of the rotational tori of the central manifold, changing the value of $u_F^0=-\Omega_0$ shifts the whole structure along the torus (see  Figure \ref{fig:theoretical_res_color} (right)  and Figure \ref{fig:inst_20000_2g_h}). 

\begin{figure}[h]

    \centering
    \includegraphics[width=0.90\textwidth]{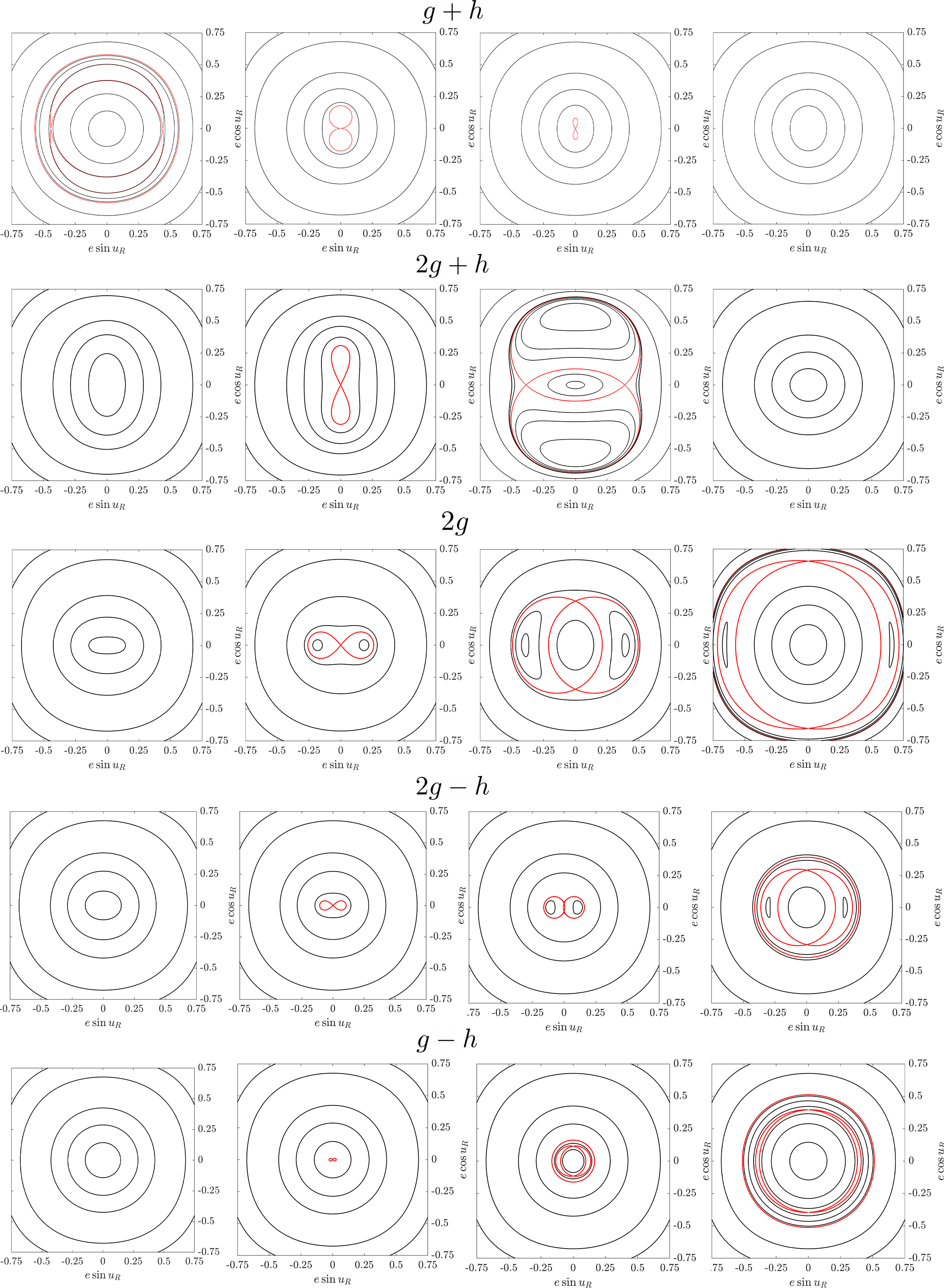}
    
    \caption{Series of bifurcations for the integrable phase portrait of inclination-only dependent lunisolar resonances at $20000$ km. 
    From left to right, the value of the inclination increases:
    $g+h: i = 44^\circ, 46.54^\circ, 46.7^\circ, 46.8^\circ; \;$
    $2g+h: i = 55.7^\circ, 56.1^\circ, 56.9^\circ, 59.5^\circ; \;$
    $2g: i = 63.2^\circ, 63.8^\circ, 65.8^\circ, 70.6^\circ; \;$
    $2g-h: i = 69.1^\circ, 69.3^\circ, 69.5^\circ, 71.7^\circ; \;$
    $g-h: i = 73.45^\circ, 73.47^\circ, 74^\circ, 79.7^\circ. \;$
    }
    
    \label{fig: bifurcations}
\end{figure}

We are now ready to study the typical bifurcations of the motion for changing values of $i$: for $i<i_1$  the point $X=0,$ $Y=0$ is stable, and we have an elliptic point, for $i \in [i_1,i_2]$ $X=0,$ $Y=0$ becomes hyperbolic and we have a figure $8$ separatrix getting bigger and bigger for increasing values of $i$. When $i>i_2$ we go back to an elliptic point in $X=0,$ $Y=0$ , but the other two hyperbolic points start to form and we have in total 5 critical points, 3 elliptic and 2 hyperbolic.
We have a "vertical" separatrix for the $2g+h$ $g+h$ and $g-h$, while we have an "horizontal" one for the resonances $2g$ and $2g-h$. For the $g+h$ resonance the evolution is reversed (see Figure \ref{fig: bifurcations}).

\section{Recovering the structure of FLI stability maps}
\label{sec: FLI Map Prediction}
In this section, we will see how we can predict the structure of the separatrices in the FLI maps using the analytic theory developed so far.
\begin{figure}[h]
\centering
  \includegraphics[width=\textwidth]{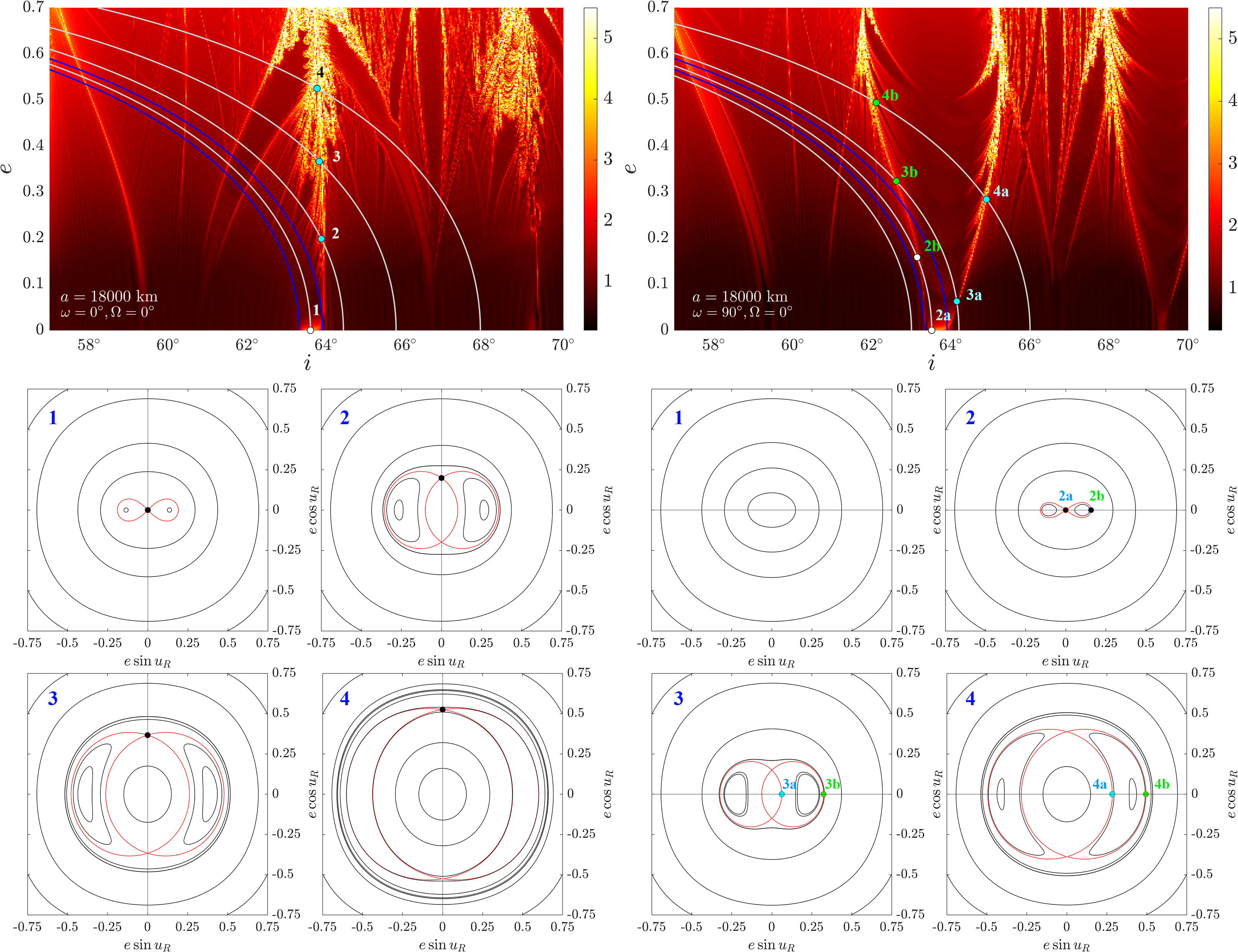}  
\caption{Analytic predictions for the vertical (left) and horizontal (right) scanning direction. Blue lines in the FLI maps correspond to the fast drift planes for $i_1$ and $i_2$.
In this plot, first four values of the inclination are chosen. For each of these values, for $e=0$, one gets an integrable phase portrait, corresponding to four values of $J_F$. 
Since $J_F$ is conserved along a fast drift plane, the same integrable phase portrait holds along the fast drift plane. 
The FLI map shows the point obtained by intersecting the separatrix of the integrable phase portrait with the scanning direction dictated by the choice of the initial angles $u_R$, which is a combination of the keplerian elements $\omega$ and $\Omega$. This point will have an eccentricity $e_p$.
Then, by intersecting the fast drift plane with the line of constant eccentricity $e_p$ one finds a point in the FLI map.}
\label{fig:Anal_Pred_18000Km}
\end{figure}
We stress that the model we constructed starts from a solution of $J_F(t)$ valid for $e=0$. Anyway, by exploiting the fact that there are planes in the phase space where $J_F$ is constant, which we call \textit{fast drift planes}, we will be able to provide analytic predictions which are accurate even for high values of the eccentricity.

\subsection{The Procedure}
We will start by providing all the procedure details in the particular case of the $2g$ resonance.
The same steps can be used for any inclination-only dependent lunisolar resonance.

We first fix a value for $a$ and $u_F^0$ .
We then find the integrable model  $\bar{H}\left(X, Y ; J_{F}^{0}, u_{F}^{0}, a\right)$ following the method explained in the previous section. 
Next, we compute the two values of the inclination $i_1$ and $i_2$ limiting the instability region for circular orbits.
Figure \ref{fig:Anal_Pred_18000Km} shows the results for $a=18000$ $ \si{\km}$, $u_F^0=0 \degree$. 
We consider a series of phase portraits showing all bifurcations occurring in this region of interest.

The FLI map shows the separatrices that we see in the plane $X$, $Y$ (or in the normalized variables $e \cos u_R$, $e \sin u_R$), and the value of the angle $u_R$ changes the "scanning" direction with which we see these separatrices.
We recall that for the $2g$ resonance $u_{R} =p-q$. 
Since we took $\omega_0 = 0$, $\Omega_0 = 0$, we have $u_R=0$, thus we are scanning the phase portrait of the integrable model vertically.

We proceed in the following way (see Figure \ref{fig:Anal_Pred_18000Km} (left)).
\begin{itemize}
\item First we sample  on the line $e=0$ some points with inclination between $i_1$ and $i_2$. 
\item For each of these points we will have in the $(e \cos u_R$, $e \sin u_R)$ plane a phase portrait with a figure $8$ separatrix.
\item We seek the value of eccentricity for which the line of our scanning direction (in our case the vertical axis) intersects the separatrix. 
\item We follow the fast drift plane for the initial inclination considered up to the value of eccentricity found in the previous step.
\end{itemize}

In conclusion, between $i_1$ and $i_2$ we will just see a horizontal line at $e=0$, corresponding to the hyperbolic point at the origin, while after $i_2$ we see a line corresponding to the positions of the hyperbolic point of the "second" separatrix.

\begin{figure}
    \centering
      \includegraphics[width=\textwidth]{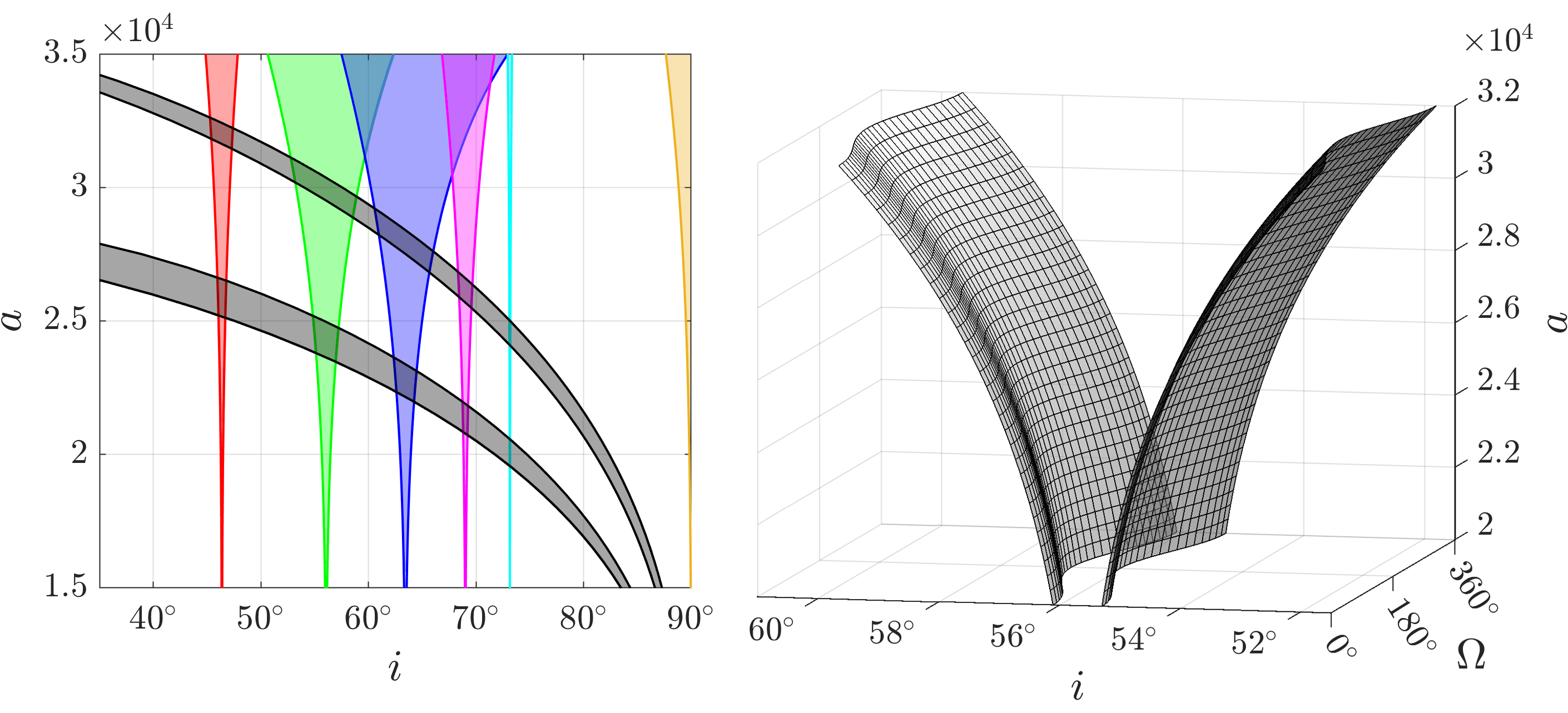}  
     \caption{(Left) Theoretical amplitude of the resonances considered in this paper as a function of the altitude.
     The resonances are: $g+h$ (red), $2g+h$ (green), $2g$ (blue), $2g-h$ (magenta), $g-h$ (cyan), and in gray the $\Omega-\Omega_L$ (lower) and $2 \Omega-\Omega_L$ (higher).
     The amplitude for those last two lunar resonances is computed for circular orbits.
     The theoretical extension for the inclination-dependent resonances has been computed using the "Breiter" model. 
     (Right) Analytical 3D representation of the $2g+h$ resonance.
     }
    \label{fig:theoretical_res_color}
\end{figure}

On the other hand, if we set initially  $\omega_0=90 \degree$, still with $u_F^0=0 \degree$, we will have  $u_R=-90 \degree$, so our scanning direction becomes horizontal.
As a consequence, besides the segment corresponding to the origin, between $i_1$ and $i_2$ we also find a curve in the $(i-e)$ plane, and after $i_2$ we intersect the "second" separatrix in two points. Thus, we see two curves in the FLI map (see Figure \ref{fig:Anal_Pred_18000Km} right).

Summing up, according to the choice of $u_F^0$ the hyperbolic region shifts along the tori corresponding to the two values $i_1$ and $i_2$, and the scanning direction changes with $u_R$. Hence the FLI map will change accordingly.

This approach works well when the true dynamics is not too far from the integrable one. 

Anyway, even when the inclination-only dependent lunisolar resonance at hand interacts with another resonance, we are able to predict the extension of the hyperbolic region for circular orbits, and thus the width of the resonance (see for example Figure 5 and Figure 6 of \cite{daquin2021deep}).

A commentary for altitudes between $17000$ km and $32000$ km for all resonances is given in the appendix.

\section{Eccentricity growth}
\label{sec:Eccentricity growth}

\begin{figure}[b]
    \centering
    \includegraphics[width=\textwidth]{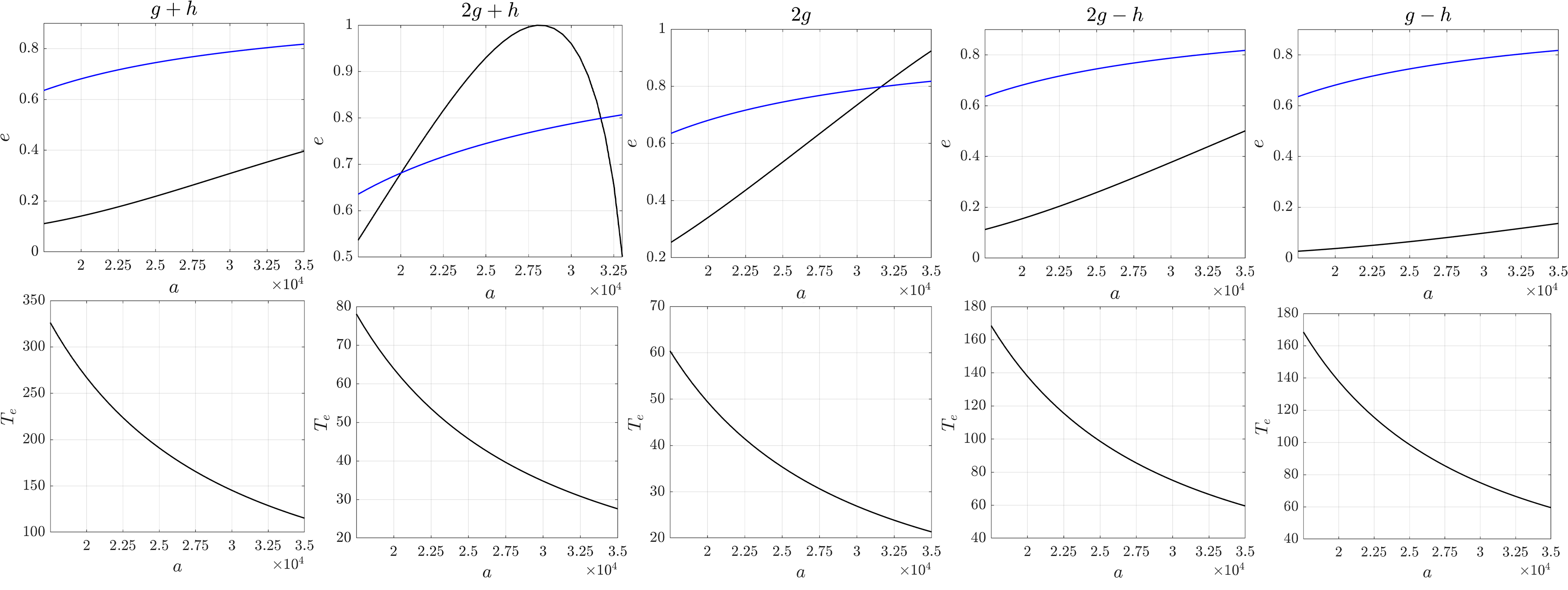}
    \caption{Top: The maximum eccentricity reachable for circular orbits close to an inclination dependent resonance as a function of the semimajor axis $17000 \si{km} \leq a \leq 32000 \si{km}$ (black line) in comparison with the eccentricity  for which the orbit’s perigee $a (1-e)$ is equal to the Earth’s radius (blue line).
    These plots show that a disposal strategy which exploits the eccentricity growth due to diffusive transport is essentially  possible only close to the $2g+h$ resonance or close to the $2g$ one for high altitudes.
    Bottom: The $e$-folding time $T_e$ as a function of the semimajor axis ranging from $17500 \si{km}$ to  $35000 \si{km}$.}
    \label{fig:emax_efolding}
\end{figure}

Since navigation satellites have almost circular orbits, we are interested in finding out which is the maximum eccentricity $e_{\max}$ that an object starting with an almost zero eccentricity could reach moving along the unstable manifold of the center manifold.
When circular orbits are unstable, we have a figure-eight separatrix in the $X-Y$ plane. In order to find $e_{\max}$,  we need to compute the uppermost point of such separatrix, which is reached at the end of the instability domain $i_2$ for all resonances but the $g+h$ one, where such maximum is reached at $i_1$.
For simplicity, we will provide these results starting from a "Breiter-like" Hamiltonian:
\begin{align}
\begin{split}
    H_{BR}(&X, Y, J_F; a) =  \; \omega_F J_F + \alpha^2 J_F^2+R_{20} X^2+ \\
    & R_{02} Y^2+R_{40} X^4+ R_{04} Y^4 + M_{140} J_F X^4+ M_{104} J_F Y^4 +  \\
	& R_{60} X^6 + R_{06} Y^6+M_{120} J_F X^2+M_{102} J_F Y^2+ \\
	& M_{220}J_F^2 X^2 + M_{202} J_F^2 Y^2 + \\
	& R_{22} X^2 Y^2 + R_{42} X^4 Y^2 + R_{24} X^2 Y^4.
\end{split}
\end{align}

We can proceed in the following way. Similarly to \ref{sec: bifurcations}, the Hessian of $H_{BR}$ for the variables $X, Y$ is diagonal, so it's eigenvalues are just
\begin{align}
\begin{split}
    \lambda_1 &= \frac{\partial^2 H_{BR}}{\partial X^2}\bigg\rvert_{X=0, Y=0} = 2 ( R_{20} +M_{120} J_F + M_{220} J_F^2), \\
\lambda_2 &= \frac{\partial^2 H_{BR}}{\partial Y^2}\bigg\rvert_{X=0, Y=0} = 2 ( R_{02} +M_{102} J_F + M_{202} J_F^2).
\end{split}
\end{align}

Consider their product $f_e(J_F; a) = \lambda_1 \lambda_2$. Let $J_F^{(1)}$, $J_F^{(2)}$ be the solutions to $f_e = 0$. They corresponds to the inclination bound within which the center orbits are unstable:
\begin{equation}
   i_{1,2} = \arccos \left(  \cos i_\star - \frac{J_F^{(1,2)}}{L} \right). 
\end{equation}

Now, we can find the highest point of the figure-8 separatrix $Y_{max}$ by solving

\begin{equation}
Y_{max} = 
    \begin{cases}
      H_{BR}\bigg\rvert_{X=0, J_F=J_F^{(2)}} = H_{BR}\bigg\rvert_{X=0, Y=0} \\
      Y > 0 
    \end{cases}
\end{equation}
(or a similar system for horizontal figure-8 separatrices).

Now, the maximum eccentricity reachable for circular orbits close to an inclination-only dependent lunisolar resonance is given by
\begin{equation}
    e_{max} = \sqrt{1- \left(1-\frac{Y_{max}^2}{2 L} \right)}.
\end{equation}

The figure-8 separatrix associated to the unstable origin reaches its maximum size at $i=i_{\star}$. So, from the fourth-order truncation of $H_{BR}$ in $X$ and $Y$, we have $Y_{max}^2 = R_{02}/R_{04}$.
These steps bring to equations (46) and (47) of \cite{daquin2021deep}, which give a simplified formula for $e_{max}$ in the case of the $2g+h$ resonance.
An analogous formula can be given for the $2g$ resonance:

\begin{equation}
    e_{\max }=\sqrt{1-\left(1-\left(1.15 \times 10^{-4}\right)\left(\frac{a}{R_{\mathrm{E}}}\right)^{5}\right)^{2}}.
\end{equation}

Numerical values for $e_{max}$ compared against the re-entry eccentricity can be found in the top row of Figure \ref{fig:emax_efolding}.

From these numerical values, we conclude that a disposal strategy that exploits the eccentricity growth due to diffusive transport is essentially possible only close to the $2g+h$ resonance or close to the $2g$ one for high altitudes.

In addition, the growth of the eccentricity along the unstable manifold of the central manifold is exponential in time, and it is possible to estimate the $e$-folding time $T_e$ as the inverse of the positive real eigenvalue (see the bottom of Figure \ref{fig:emax_efolding}).

\begin{figure}
    \centering
    \includegraphics[width = \textwidth]{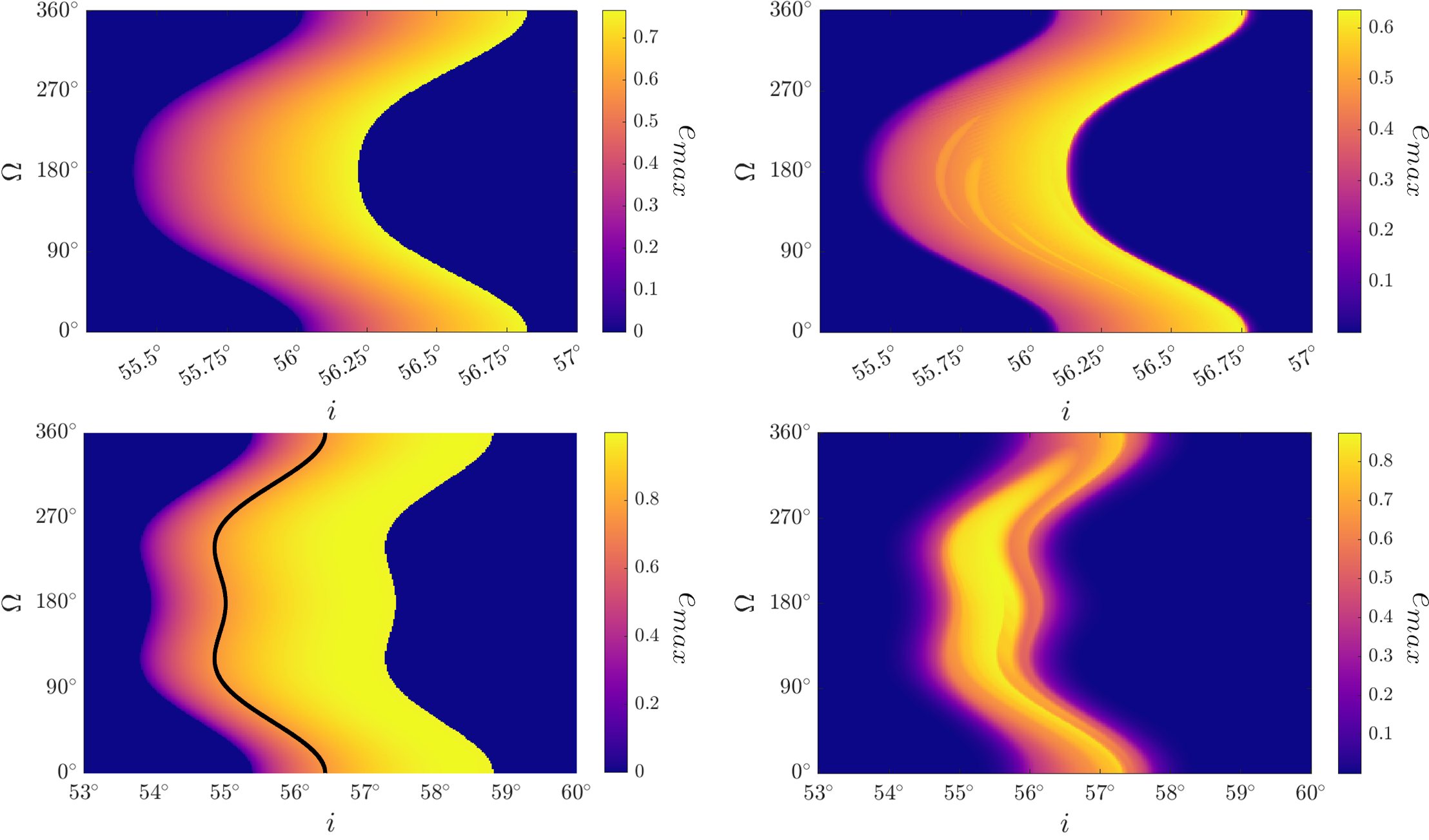}
    \caption{Computation of the maximum eccentricity reachable along the $2g+h$ resonance for $a = 20000$ km and $27000$ km in the $(i, \Omega)$ plane. Left: analytical computation, using the procedure described on section \ref{sec:Eccentricity growth} for the computation of $e_{max}$, using instead of $H_{BR}$ the integrable Hamiltonian computed in Section \ref{sec: Analytic Theory} with the $LL$ method.
    The values for $i_1$ and $i_2$ are shifted along the corresponding tori of the Center Manifold as the initial value of the angle $\Omega$ changes. The values of $e_{max}$ increase with the size of the figure-8 separatrix of the resonance, and the maximum value is obtained on the torus corresponding to $i_2$.
    Right: numerical computation up to $1000$ years (at $20000$ km) and $500$ years (at $27000$ km) on a $300$x$300$ grid in the $(i, \Omega)$. 
    For every point of the grid $(i^*,\Omega*)$, $e_{max}$ has been computed by propagating a grid $60$x$60$ of initial conditions for the remaining keplerian elements $e \in [10^{-5}, 10^{-4}]$ and $\omega \in [0, 2 \pi]$; then, $e_{max}$ for the point $(i^*,\Omega*)$ is given by the maximum eccentricity reached over the $(e,\omega)$ grid.
    In the "ordered" domain of $20000$ km we have a good agreement between the two plots. At $27000$ km the value of $e_{max}$ depends on the structure of the manifolds. In addition, the integrable model gives values of the eccentricity greater than the re-entry one (whose value is plotted as the black contour line).
    See Figure 5 and Figure 6 of \cite{daquin2021deep} for the FLI maps at these altitudes.
   }
    \label{fig:emax}
\end{figure}

The Hamiltonian $H_{BR}$ could be regarded as an average over the angles of the motion.
If we follow this same procedure using the more realistic model obtained in Section \ref{sec: Analytic Theory}  we would get almost the same value for $e_{\max}$, but the location of such point would be shifted in inclination with varying values of $\Omega$.
This explains the results obtained in previous works, such as \cite{alessi2016numerical}, where the simulations were carried out for three different values of inclination, a nominal one and the plus or minus one degree. 
The different behaviors found for the possible eccentricity growth are explained in Figure \ref{fig:emax}. 
Here on the left the maximum eccentricity $e_{max}$ reachable along the $2g+h$ resonance is computed analytically, while on the right the same value is computed numerically by propagating a grid of initial conditions.
From Figure \ref{fig:emax} we can conclude  that the theory works  
well in the non-chaotic domain (as expected, since the theory 
is based on an integrable approximation). 
On the other hand, in domains 
of strong chaos, we actually see that the theory overestimates the possible 
eccentricity growth. 
This is understandable, since the theoretical 
separatrices of the integrable model (which, in reality, are destroyed 
in the strong chaos regime), keep growing as the initial inclination 
increases, even after the eccentricity for which we would have atmospheric re-entry (shown as the green contour line in Figure \ref{fig:emax}). 
Nevertheless, the theoretical model keeps explaining 
the dependence of the eccentricity growth on the initial phase $\Omega$. 
This is due to (and consistent with) the mechanism explained in Section 
\ref{sec: Analytic Theory}, i.e., the shift of the curve representing the intersection of the plane 
of fast drift with the plane $(i, e)$ as $\Omega$ changes, due to the 
structure of the invariant tori of the associated center manifold of 
circular orbits. 

\section{Conclusions}

We have seen how the integrable model developed in Section \ref{sec: Analytic Theory} can be used to understand the structure of inclination-only dependent lunisolar resonances and to compute the correct extension of their separatrices (as shown in Section \ref{sec: FLI Map Prediction}).  Also, the model can be used to compute the limiting values $i_1$ and $i_2$ within which circular orbits are unstable, and thus where the central manifold becomes normally hyperbolic. This can be used to estimate the width in inclination of a given resonance for any semi-major axis $a$. 
In the appendix, we provide a report for every resonance.
We built our integrable model assuming a single resonance. Because of this, the prediction of the separatrices is accurate far from the intersections with the $\Omega-\Omega_L$ and $2 \Omega- \Omega_L$ resonances, but in any case, the prediction of the width of the resonance is always quite accurate.
Moreover, we have seen how our model is able to explain the dependence of the FLI maps from the phases of the angles $\omega, \Omega$ and $\Omega_L$: given a particular resonance, the combination of the angles $\omega$ and $\Omega$ gives the value of the angle $u_R$ which defines the direction in which the integrable phase portrait is seen. Then,  the initial phase of $\Omega_L$ affects the shape of the center manifold (the whole phase portrait is moved up or down), and the angle $\Omega$ defines where we are on a particular torus of the center manifold. So in this way the instability region where the central manifold becomes normally hyperbolic is shifted along the two limiting tori.
 Such prediction is show in the left of Figure \ref{fig:theoretical_res_color}.  
 In addition, we have seen in Figure \ref{fig:emax_efolding} that our integrable approximation shows that eccentricity-growth re-entry is possible for the $2g+h$ resonance and for the $2g$ one when the altitude is high enough.

\bmhead{Acknowledgments}
We acknowledge the support of the Marie Curie Initial Training Network Stardust-R, grant agreement
Number 813644 under the H2020 research and innovation program: https://doi.org/10.3030/813644.

\clearpage
\bigskip

\begin{appendices}

    \begin{figure}
        \centering
         \includegraphics[width=\textwidth]{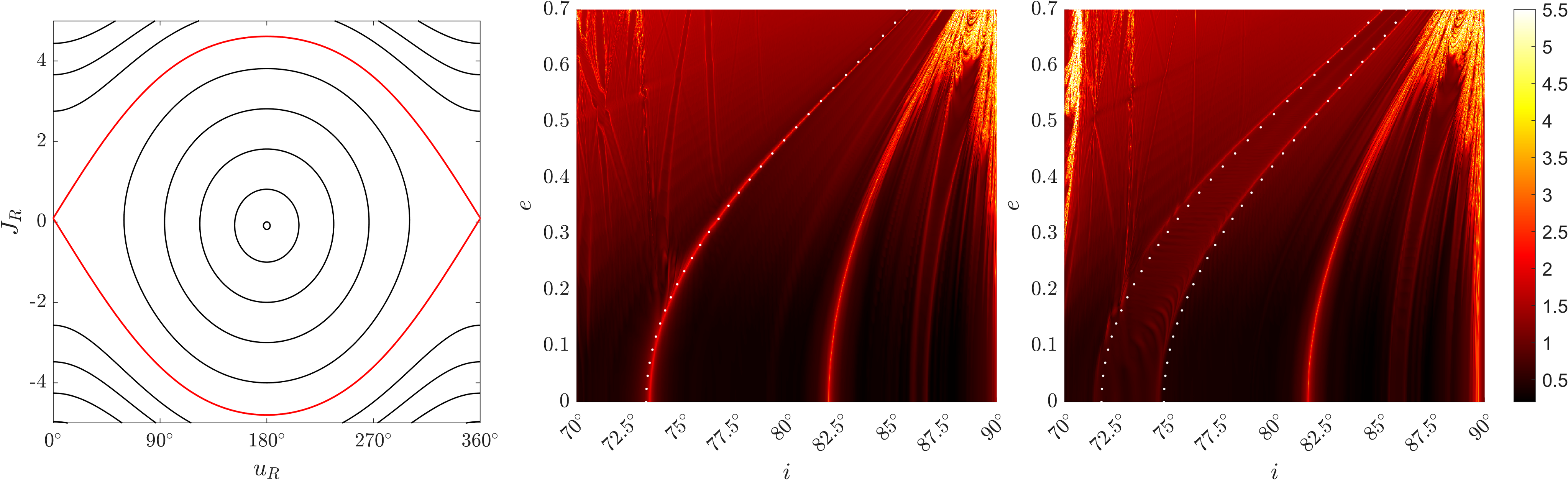}  
         \caption{ Resonance $\Omega-\Omega_L$. Altitude: $20000$ $ \si{\km}$. 
         (Left) Phase portrait at the point $(i,e)=(73.31 \degree, 0.3)$. 
         (Center) Analytic prediction over FLI map for $\Omega=0\degree$ and (Right) $\Omega=180 \degree$.
         }
        \label{fig:OM-OML}
        \end{figure}
        \appendix
        
        \section{The resonances $\Omega-\Omega_L, 2\Omega-\Omega_L$}
        
        As highlighted in the main body of this paper, the two resonances $\Omega-\Omega_L$ and  $2\Omega-\Omega_L$ have a key role in shaping the dynamics in the MEO region, and they belong to the first fundamental model of resonance.
        In this section, we will study them in greater detail.
        
        To obtain a model for these resonances, we perform
        the following steps.
        We first isolate from the Hamiltonian \eqref{HSEC} just the terms containing $\Omega-\Omega_L$ or  $2\Omega-2\Omega_L$, and then we express the result in modified Delaunay coordinates.
        The center of the resonance is $\mathcal{C} = \left\lbrace (e,i): \nu_{\Omega}=\nu_{\Omega_L} \right\rbrace$. 
        We fix a point $\left( P^\star, Q^\star \right)\in \mathcal{C}$. 
        We perform a Taylor expansion around $\left( P^\star, Q^\star \right)$.
        Next, we apply the canonical transformation generated by $S=(q+\Omega_L) J_R+p J_1+\Omega_L J_2$.
        In the resulting Hamiltonian, we would have $J_R (\nu_\Omega+\nu_{\Omega_L})$, so the linear term vanishes.
        Also, after the canonical transformation, we will not have the angles $u_1$ and $u_2$, so $J_1$ and $J_2$ are constants of motions, and we set them to 0.
        Finally, we obtained an Hamiltonian $H_F$ function of just $u_R$ and $J_R$, which has the form
        $$H_F=C+\alpha J_R^2-\gamma_1 \cos u_R +\gamma_2 \cos 2u_R+ \delta J_R \cos u_R + \dots$$
        for suitable $C, \alpha, \gamma_1, \gamma_2,\delta \in \mathbb{R}$, and its phase portrait is equivalent to the one of the pendulum (see Figure \ref{fig:OM-OML}).
        Again, in the FLI map we see the separatrix of such resonance, thus when $u_R=0$ we will just see the hyperbolic point, while for $u_R=180 \degree$ we will have two lines corresponding to the points obtained by intersecting a vertical line in $u_R=180 \degree$ with the separatrix.
        By repeating this step for a suitable number of points in the center of the resonance $\mathcal{C}$ we can plot an analytic prediction of the extension of the resonance over the FLI map (see Figure \ref{fig:OM-OML}).

        \section{Analytic Predictions}
        \label{app:FLI_MAPS}
        
\begin{figure*}
            \centering
            \includegraphics[width = 1 \textwidth]{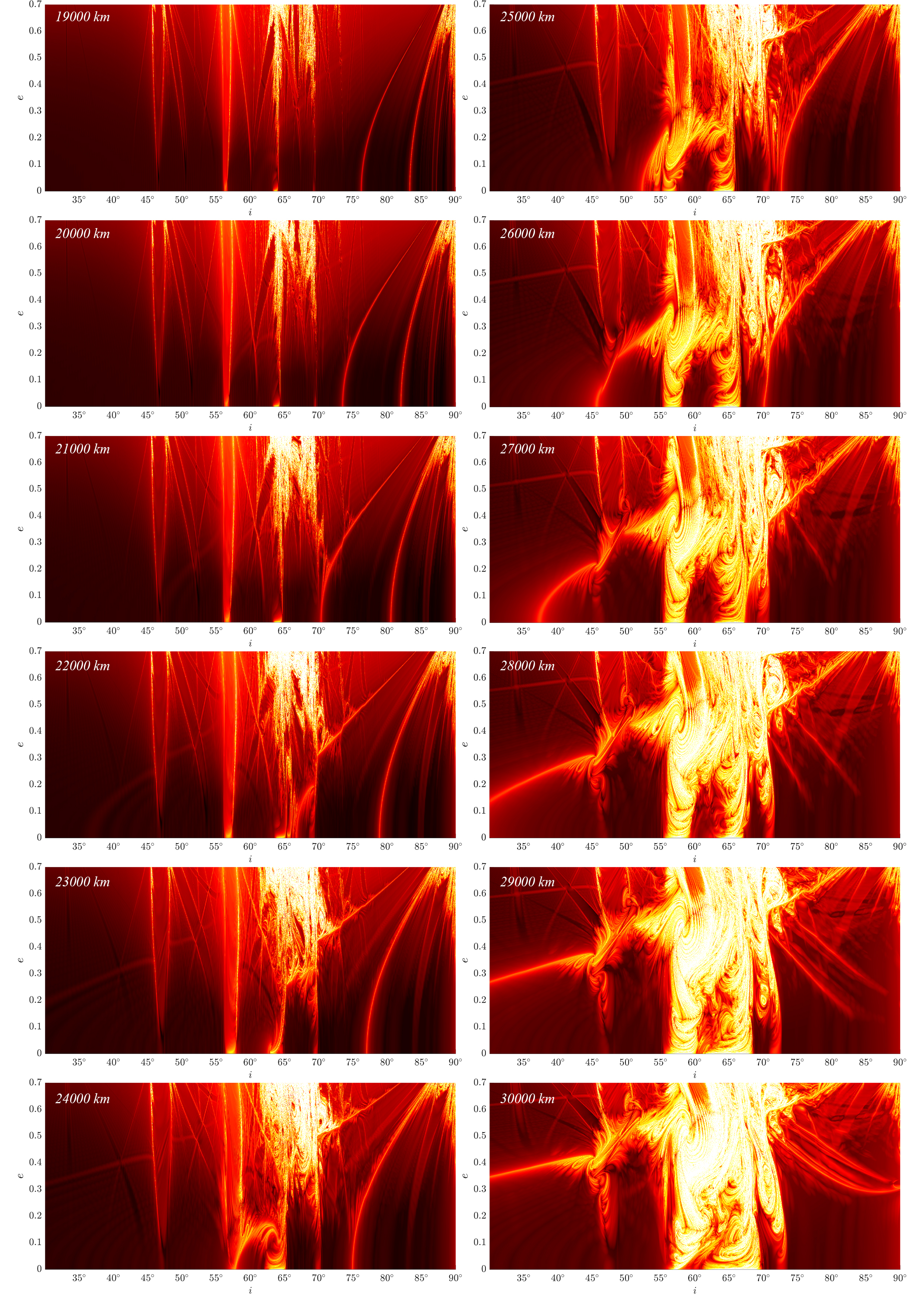}
            \caption{FLI cartography in the $e \in [0, 0.7]$, $i \in [30^\circ, 90^\circ]$ plane for $\omega = 0^\circ, \Omega = 0^\circ$, $a \in [19000 \text{ km}, 30000 \text{ km}]$, }
            \label{fig: FLI_grid_OM0}
        \end{figure*}

\begin{figure*}
            \centering
            \includegraphics[width = \textwidth]{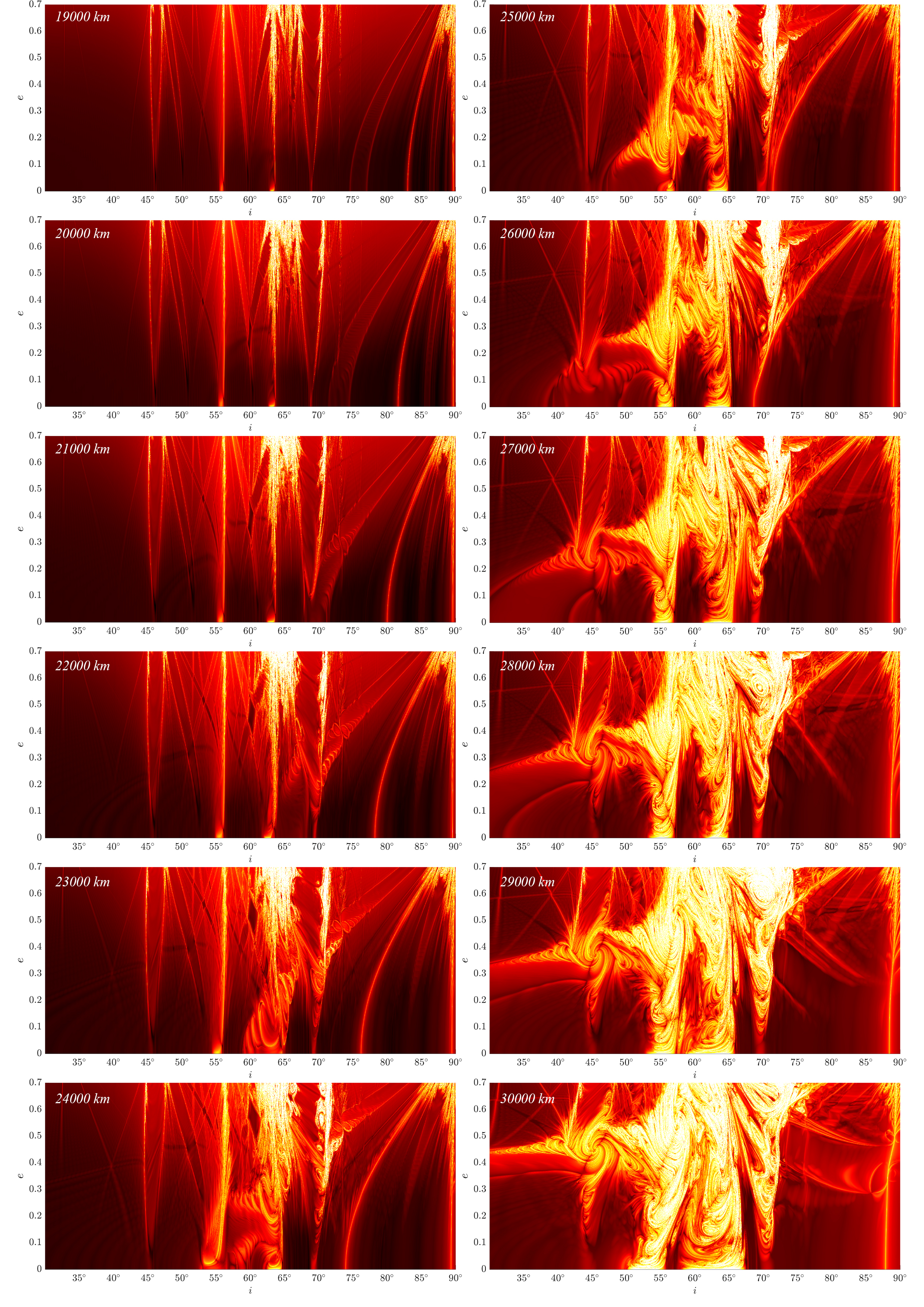}
            \caption{FLI cartography in the $e \in [0, 0.7]$, $i \in [30^\circ, 90^\circ]$ plane for $\omega = 0^\circ, \Omega = 180^\circ$, $a \in [19000 \text{ km}, 30000 \text{ km}]$, }
            \label{fig: FLI_grid_OM180}
        \end{figure*}

        In this section, we will provide a detailed analysis of the applicability of the method discussed in Sections \ref{sec: Analytic Theory} - \ref{sec: FLI Map Prediction} for the lowest order inclination dependent resonances.
        Here we provide a commentary for every altitude and resonance, highlighting which of the approaches described in this paper works best in that particular case and which lunar resonance intersects the domain if there is one. 
        You can compare this against the FLI cartography in Figure \ref{fig: FLI_grid_OM0} and Figure \ref{fig: FLI_grid_OM180}
        
        \subsection{$g+h$ resonance}
        \begin{itemize}
        \item Until  $22000$ km, the "linear" approach works better.
        \item Between $22000$ km and $25000$ km, the "pendulum" approach works better.
        \item Above $26000$ km, the sweeping of the $\Omega-\Omega_L$ resonance prevents us from obtaining good predictions.
        \end{itemize}

        \subsection{$2g+h$ resonance}
        
        \begin{itemize}
        \item Below $23000$ km, everything works fine with the "linear" approach.
        \item Between $24000$ km and $25000$ km the $2g+h$ resonance interacts with the $\Omega-\Omega_L$.
        This leads to a loss in accuracy of the analytic prediction, especially for $\Omega=180 \degree$.
        \item Between $26000$ km and $28000$ km we are in a domain where the $\Omega-\Omega_L$ affects the resonance for higher and higher values of the eccentricity, while for lower ones the chaotic region of the $2g+h$ resonance is surrounded by a stable region.
         For $\Omega=0$ the prediction of the first separatrix still gives some information about the development of the resonance. 
        \item Above $29000$ km we are in the domain of the $2 \Omega-\Omega_L$ resonance. Moreover, due to the increasing value of the Laplace plane, the domain of the $2g+h$ resonance starts intersecting the $2g$ resonance. 
        \end{itemize}
        
        \begin{outline}
        
         \1 From $17000$ km to $20000$ km we start to see the influence of the $\Omega-\Omega_L$ resonance, and we need to use the \textit{LL}  approach. At $17000$ km we could also use the $L$ one. 
         \1 At $21000$ km, we have the intersection with the $\Omega-\Omega_L$ resonance.
         \1 At $22000$ km, the dynamics is still shaped by the interaction with the $\Omega-\Omega_L$ resonance, and because of this, the model predictions are not very accurate.
         \1 From $23000$ km to $24000$ km the predictions are back to be fairly accurate. Since we are in a transition zone slightly influenced by both the   $\Omega-\Omega_L$ and the $2 \Omega-\Omega_L$ resonance, the approach that works better is the $LL$ one. 
         \1 At $25000$ km, we start to really see the influence of the $2 \Omega-\Omega_L$ resonance. 
          \1 At $26000$ km, we have the intersection with the $2 \Omega-\Omega_L$ resonance. Every approach fails and it is not possible to provide accurate predictions.
        \1    From $27000$ km to $28000$ km, we start to be able to provide some insights with our analytical method, but we begin to see the influence of the $2g$ resonance starting to intersect the domain of the $2g+h$ one. In this case, the better estimate is given by the $L$ method, with the tori of the $P$ one.
        \1 Above $29000$ km, the intersection of the $2g+h$ resonances with the $2 \Omega-\Omega_L$ and the $2g$ ones makes the system too much chaotic to provide analytic predictions.
        
        \end{outline}
        
        \subsection{$2g$ resonance}
        
        \begin{outline}
        \1 Until $19000$ km,  the "linear" approach \textit{L} works better.
        \1 Between $20000$ $\si{\km}$ and $24000$ $\si{\km}$, the dynamics is influenced by the $\Omega-\Omega_L$ resonance.
        \2 From $20000$ $\si{\km}$  when we begin to see the influence of the $\Omega-\Omega_L$ resonance, the "pendulum" approach starts to work better for low eccentricities, but this approach fails for higher ones.
        \2 For $22000$ km and $23000$ km, we are in a chaotic domain close to the $\Omega-\Omega_L$ resonance. 
        \2 Between  $23000$ km  and  $24000$ km, the influence of the $\Omega-\Omega_L$ resonance is still strong. 
        It is best to use the predictions of the linear \textit{L} method with the tori of the \textit{LL} one. 
        \1 At  $25000$ $\si{\km}$  we are in a transition zone where is dynamics is influenced by both the $\Omega-\Omega_L$ and the $2 \Omega-\Omega_L$ resonance. The approach that works better is the \textit{LL} one.
        \1 From $26000$ $\si{\km}$ to $32000$ $\si{\km}$ we are in the domain of the $2 \Omega-\Omega_L$ resonance. 
        \2 At  $28000$ km we are very close to the $2 \Omega-\Omega_L$ resonance. The linear approach fails, and so we use the \textit{P} one. Being so close to the $2 \Omega-\Omega_L$ resonance, our predictions can't be accurate, but nevertheless, the second separatrix in the $\omega=90 \degree$ gives a sense for the extension of the chaotic region.
        \2 Above  $29000$ km both the \textit{LL} and the \textit{P} approaches fail. 
        Also, the domain of the $2g+h$ and of the $2g$ resonance start to overlap.
        The best we can do is to use the predictions of the \textit{L} method with the tori of the \textit{P} one.
        \2 At $31000$ km and $32000$ km  the domain of the $2g+h$ and of the $2g$ resonance overlap. Thus the region is very chaotic and the predictions cannot be accurate.
        \end{outline}
        
        \subsection{$2g-h$ resonance}
        
        \begin{outline}
        
         \1 From $17000$ $ \si{\km}$ to $20000$ $ \si{\km}$ we start to see the influence of the $\Omega-\Omega_L$ resonance, and we need to use the \textit{LL}  approach. At $17000$ $ \si{\km}$ we could also use the $L$ one. The behavior of the pendulum method \textit{P} is the same as in the previous resonance $2g$: it works very well for low eccentricities but fails for higher values. 
         \1 At $21000$ $ \si{\km}$ we have the intersection with the $\Omega-\Omega_L$ resonance.
         \1 At $22000$ $ \si{\km}$ the dynamics is still shaped by the interaction with the $\Omega-\Omega_L$ resonance, and because of this, the model predictions are not very accurate.
         \1 From $23000$ $ \si{\km}$ to $24000$ $ \si{\km}$ the predictions are back to be fairly accurate. Since we are in a transition zone slightly influenced by both the   $\Omega-\Omega_L$ and the $2 \Omega-\Omega_L$ resonance, the approach that works better is the $LL$ one. 
         \1 At $25000$ $ \si{\km}$ we start to really see the influence of the $2 \Omega-\Omega_L$ resonance. 
          \1 At $26000$ $ \si{\km}$ we have the intersection with the $2 \Omega-\Omega_L$ resonance. Every approach fails and it is not possible to provide accurate predictions.
        
        \end{outline}

        \subsection{$g-h$ resonance}
        
        \begin{outline}
        
         \1 From $17000$ $ \si{\km}$ to $19000$ $ \si{\km}$, predictions work well with the  \textit{LL}  approach. This region is slightly influenced by the $\Omega-\Omega_L$ resonance.
         \1 At $20000$ $ \si{\km}$, we have the intersection with the $\Omega-\Omega_L$ resonance. We can't provide predictions also at $21000$ $ \si{\km}$. 
         \1 At $22000$ $ \si{\km}$,  the predictions are back to being fairly accurate. Since we are in a transition zone slightly influenced by both the   $\Omega-\Omega_L$ and the $2 \Omega-\Omega_L$ resonance, the approach that works better is the $LL$ one. 
         \1 From $23000$ $ \si{\km}$, we start to really see the influence of the $2 \Omega-\Omega_L$ resonance. 
          \1 At $25000$ $ \si{\km}$, we have the intersection with the $2 \Omega-\Omega_L$ resonance. Also, we start to see the merging of the $g-h$ resonance with the $2g-h$ one.
        \1    Above $26000$ $ \si{\km}$, the intersections with the other resonances (mainly the $2g-h$ and the $2 \Omega-\Omega_L$ ones) destroy completely the structure provided by the integrable approximation.
        
        \end{outline}
        
        \newpage
        \clearpage
        
        \section{Coefficients}
        
        In this section, we report the explicit expressions of coefficients of the resonant Hamiltonian in the case of each of the resonances of Table \ref{main_res}.
        
        We set the unit of time to be one synodic day $d$, and the unit of length to be the Earth radius, $R_E$.
        The numerical values of the constants involved are reported in Table \ref{tab: num_constants}.
        Consider an inclination dependent resonance $\sigma=2 \omega + m_2 \Omega$, located at $i_\star$.
        The coefficients of the terms independent of the variables $X$ and $Y$ do not depend on $m_2$. Their analytic formula is given in Table \ref{tab: CM}. 
        We abbreviate $\sin \alpha$ and $\cos \alpha$ with $s \alpha$ and $c \alpha$. 
        Notice that the coefficient of $J_F^2$ does not depend from $i_\star$. 
        The numerical value is
        $$\left( \dfrac{-8.120\times 10^{-4}}{(\frac{a}{R_E})^{^4}}-(2.329\times10^{-8})\frac{a}{R_E}\right) \; \frac{1}{R_E^2}$$
         for every resonance.
         
         A general analytic expression for the coefficients of the terms depending from $X$ or $Y$ could be given, but it is quite cumbersome. Thus, for these coefficients we provide the numerical values.
        We recall that $X$ and $Y$ have the dimension of a square root of an action, and that an action has the dimension of energy times time.
         
        \begin{table}[b]
            \centering
            \small
            \caption{Numerical values of the physical constants.}
            \begin{adjustbox}{width=\textwidth}
            {\renewcommand{\arraystretch}{1.4}
            \begin{tabular}{|c|c|c|}
            \Xhline{1.15pt}  
            Symbol  & Description & Numerical Value\\
            \Xhline{1.15pt}  
            $R_E$  & Earth Radius & $6378.1363$ $\si{km}$ \\ 
            $\mu_E$  & Standard Gravitational Parameter $G M_E$ of the Earth & $11467.9 \;\frac{R_E^3}{d^2}$ \\ 
            $\mu_L$  & Standard Gravitational Parameter $G M_L$ of the Moon & $141.056 \;\frac{R_E^3}{d^2}$ \\
            $\mu_S$  & Standard Gravitational Parameter $G M_S$ of the Sun & $3.81819 \times 10^9 \;\frac{R_E^3}{d^2}$\\
            $r_L$  & Radius of the Moon & $60.2303 \; R_E$\\
            $r_S$  & Radius of the Sun & $23451.9 \; R_E$\\
            $i_E$  & Obliquity of the Ecliptic & $23.4392911 \degree$\\
            $i_L$  & Mean Inclination of the Lunar Orbit  & $5.15 \degree$\\
            $J_2$  & Earth zonal harmonic  & $1.08263 \times 10^{-3}$\\
            \Xhline{1.15pt}  
            \end{tabular}}
        \end{adjustbox}
        \label{tab: num_constants}
        \end{table}

        \begin{table}
            \centering
            \caption{Coefficients of the main terms of $H_{CM}$.}
            \begin{adjustbox}{width=\textwidth}
                {\renewcommand{\arraystretch}{2.5}
            \begin{tabular}{|c|c|}
            \Xhline{1.15pt}  
            Term   & Coefficient\\
            \Xhline{1.15pt}  
            $J_F$ &$ \omega_F = \frac{3 ci_\star J_2 \sqrt{ \mu_E } R_E^2 }{2 a^{\sfrac{7}{2}} }+\frac{3 ci_\star a^{\sfrac{3}{2}} \mu_L \left(3 si_E^2-2\right) \left(3 si_L^2-2\right)}{16 \sqrt{\mu_E} r_L^3}+\frac{3 ci_\star a^{\sfrac{3}{2}} \mu_S \left(2-3 si_E^2\right)}{8\sqrt{\mu_E}  r_S^3}$ \\ 
            \rule{0pt}{2ex} 
            $J_F^2$ &$ \alpha = -\frac{3 J_2 R_E^2}{4 a^4}-\frac{3 a \mu_L\left(3 si_E^2-2\right) \left(3 si_L^2-2\right)}{32\mu_E r_L^3}+\frac{3 a \mu_S\left(3
               si_E^2-2\right)}{16\mu_E r_S^3}$ \\ 
            \rule{0pt}{2ex} 
            $\cos u_F$  & $C_1 =  \frac{3ci_\star ci_E a^2  \mu_L si_\star si_E \left(3si_L^2-2\right)}{8  r_L^3}-\frac{3ci_\star ci_E a^2 \mu_S si_\star si_E}{4  r_S^3}$\\
            \rule{0pt}{2ex} 
             $\cos 2u_F$ & $C_2 = \frac{3 a^2 \mu_L si_\star^2 si_E^2 \left(3 si_L^2-2\right)}{32 r_L^3}-\frac{3 a^2
               ci_\star si_\star^2 si_E^2}{16 r_S^3}$\\
            \rule{0pt}{2ex} 
             $\cos(2 u_F+\Omega_L)$ & $D_{21} = -\frac{3 a^2 (1+ci_E) ci_L  \mu_L si_\star^2 si_E si_L}{16 r_L^3}$\\
            \rule{0pt}{2ex} 
            $\cos (u_F+\Omega_L)$ & $D_{11} =-\frac{3 a^2 ci_\star ci_L \mu_L si_\star si_L \left(ci_E-2 si_E^2+1\right)}{8 r_L^3}$\\
            \Xhline{1.15pt}  
            \end{tabular}}
        \end{adjustbox}
            \label{tab: CM}
        \end{table}
    
        \begin{table}
            \centering
            \caption{Coefficients of $X^2$, $Y^2$, $X^4$, $Y^4$.}
            \begin{adjustbox}{width=\textwidth}
                {\renewcommand{\arraystretch}{2.5}
            \begin{tabular}{|c|c|c|c|}
            \Xhline{1.15pt}  
             Resonance   & Coeff. of $X^2=-Y^2$             							 & Coeff. of $X^4$        & Coeff of. $Y^4$\\
            \Xhline{1.15pt}  
            $g+h$  &  \scalebox{.7}{ $\left( 9.235 \times 10^{-7}  \left(\frac{a}{R_E}\right)^{^3/_2} \right)  \frac{R_E^4}{d}  $} &  \scalebox{.7}{$\left(\dfrac{1.542\times 10^{-4}}{(\frac{a}{R_E})^4}+(6.912\times10^{-9})\frac{a}{R_E} \right)  \frac{R_E^4}{d} $} & \scalebox{.7}{$\left(\dfrac{1.542\times 10^{-4}}{(\frac{a}{R_E})^4}+(1.439\times10^{-8})\frac{a}{R_E} \right)  \frac{R_E^4}{d} $}  \\ 
            \rule{0pt}{3ex} 
             $2g+h$  & \scalebox{.7}{ $\left(-(3.857\times 10^{-6})  \left(\frac{a}{R_E}\right)^{^3/_2} \right)  \frac{R_E^4}{d}  $}  &\scalebox{.7}{$\left(\dfrac{-2.710 \times 10^{-5}}{(\frac{a}{R_E})^4}+(7.458\times10^{-9})\frac{a}{R_E} \right)  \frac{R_E^4}{d} $}      &\scalebox{.7}{$\left(\dfrac{-2.710 \times 10^{-5}}{(\frac{a}{R_E})^4}-(1.091\times10^{-8} ) \frac{a}{R_E} \right)  \frac{R_E^4}{d} $}\\ 
            \rule{0pt}{3ex} 
            $2g$  & \scalebox{.7}{ $\left((4.988\times 10^{-6})  \left(\frac{a}{R_E}\right)^{^3/_2} \right)  \frac{R_E^4}{d}  $}  &\scalebox{.7}{$\left(\dfrac{-2.030 \times 10^{-4}}{(\frac{a}{R_E})^4}-(3.493\times10^{-8})\frac{a}{R_E} \right)  \frac{R_E^4}{d} $}      &\scalebox{.7}{$\left(\dfrac{-2.030 \times 10^{-4}}{(\frac{a}{R_E})^4}+(1.164\times10^{-8})\frac{a}{R_E} \right)  \frac{R_E^4}{d} $}\\
            \rule{0pt}{3ex} 
            $2g-h$  & \scalebox{.7}{ $\left((1.767\times 10^{-6})  \left(\frac{a}{R_E}\right)^{^3/_2} \right)  \frac{R_E^4}{d}  $}   &\scalebox{.7}{$\left(\dfrac{-3.992 \times 10^{-4}}{(\frac{a}{R_E})^4}-(3.867\times10^{-8})\frac{a}{R_E} \right)  \frac{R_E^4}{d} $}      &\scalebox{.7}{$\left(\dfrac{-3.992 \times 10^{-4}}{(\frac{a}{R_E})^4}-(2.121\times10^{-9})\frac{a}{R_E} \right)  \frac{R_E^4}{d} $} \\
            \rule{0pt}{3ex} 
            $g-h$  & \scalebox{.7}{ $\left( (1.630\times 10^{-7}) \left(\frac{a}{R_E}\right)^{^3/_2} \right)  \frac{R_E^4}{d}  $}   &\scalebox{.7}{$ \left( \dfrac{-6.414 \times 10^{-4}}{(\frac{a}{R_E})^4}-(3.243\times10^{-8})\frac{a}{R_E} \right)  \frac{R_E^4}{d} $}     &\scalebox{.7}{$\left( \dfrac{-6.413 \times 10^{-4}}{(\frac{a}{R_E})^4}-(2.613\times10^{-9})\frac{a}{R_E} \right)  \frac{R_E^4}{d} $}\\
            \Xhline{1.15pt}  
            \end{tabular}}
        \end{adjustbox}
        \end{table}

        \begin{table}
            \centering
            \caption{Coefficients of $X^6$, $Y^6$.}
            \begin{adjustbox}{width=\textwidth}
                {\renewcommand{\arraystretch}{2.5}
            \begin{tabular}{|c|c|c|}
            \Xhline{1.15pt}  
            Resonance   & Coeff. of $X^6$   & Coeff. of $Y^6$\\
            \Xhline{1.15pt}  
             $g+h$        &  \scalebox{.75}{ $\left(\dfrac{2.248 \times 10^{-6}}{(\frac{a}{R_E})^{\sfrac{9}{2}}}+(1.299\times10^{-11})(\frac{a}{R_E})^{\sfrac{1}{2}}  \right) \frac{R_E^4}{d} $}  & \scalebox{.75}{ $\left(\dfrac{2.248 \times 10^{-6}}{(\frac{a}{R_E})^{^9/_2}}+(1.902\times10^{-11})(\frac{a}{R_E})^{\sfrac{1}{2}}  \right) \frac{R_E^4}{d} $} \\ 
            \rule{0pt}{3ex} 
            $2g+h$        & \scalebox{.75}{ $\left(\dfrac{-4.271 \times 10^{-7}}{(\frac{a}{R_E})^{^9/_2}}-(4.017\times10^{-12})(\frac{a}{R_E})^{\sfrac{1}{2}}  \right) \frac{R_E^4}{d} $ }  & \scalebox{.75}{$\left(\dfrac{-4.271 \times 10^{-7}}{(\frac{a}{R_E})^{\sfrac{9}{2}}}-(5.746\times10^{-12})(\frac{a}{R_E})^{\sfrac{1}{2}}  \right) \frac{R_E^4}{d} $} \\ 
            \rule{0pt}{3ex} 
            $2g$        & \scalebox{.75}{$\left(\dfrac{-3.475 \times 10^{-6}}{(\frac{a}{R_E})^{\sfrac{9}{2}}}-(1.087\times10^{-10})(\frac{a}{R_E})^{\sfrac{1}{2}}  \right) \frac{R_E^4}{d} $}   &\scalebox{.75}{$\left(\dfrac{-3.475 \times 10^{-6}}{(\frac{a}{R_E})^{\sfrac{9}{2}}} \right) \frac{R_E^4}{d}$}\\
            \rule{0pt}{3ex} 
            $2g-h$        & \scalebox{.75}{$\left(\dfrac{-7.376 \times 10^{-6}}{(\frac{a}{R_E})^{\sfrac{9}{2}}}-(1.753\times10^{-10})(\frac{a}{R_E})^{\sfrac{1}{2}}  \right) \frac{R_E^4}{d} $  } &\scalebox{.75}{$\left(\dfrac{-7.376 \times 10^{-6}}{(\frac{a}{R_E})^{\sfrac{9}{2}}}-(1.084\times10^{-10})(\frac{a}{R_E})^{\sfrac{1}{2}}  \right) \frac{R_E^4}{d} $}\\
            \rule{0pt}{3ex} 
            $g-h$        &\scalebox{.75}{ $\left(\dfrac{-1.261 \times 10^{-5}}{(\frac{a}{R_E})^{\sfrac{9}{2}}}-(2.716\times10^{-10})(\frac{a}{R_E})^{\sfrac{1}{2}}  \right) \frac{R_E^4}{d} $ }  &\scalebox{.75}{$\left(\dfrac{-1.261 \times 10^{-5}}{(\frac{a}{R_E})^{\sfrac{9}{2}} }-(2.822\times10^{-10})(\frac{a}{R_E})^{\sfrac{1}{2}}  \right) \frac{R_E^4}{d} $}\\
            \Xhline{1.15pt}  
            \end{tabular}}
        \end{adjustbox}
        \end{table}
            
            \begin{table}
            \centering
            \caption{Coefficients of $X^2 Y^2$,  $X^2 Y^4$.}
            \begin{adjustbox}{width=\textwidth}
                {\renewcommand{\arraystretch}{2.5}
            \begin{tabular}{|c|c|c|}
            \Xhline{1.15pt}  
            Resonance   & Coeff. of $X^2 Y^2$   & Coeff. of $X^2 Y^4$\\
            \Xhline{1.15pt}  
             $g+h$        & \scalebox{.75}{$\left(2.130 \times 10 ^{-8} (\frac{a}{R_E}) + \dfrac{3.093 \times 10^{-4}}{(\frac{a}{R_E})^4} \right) \frac{1}{R_E^2} $ }  & \scalebox{.75}{ $\left( \dfrac{6.743\times 10^{-6}}{(\frac{a}{R_E})^{\sfrac{9}{2}}}+(5.103\times10^{-11})(\frac{a}{R_E})^{\sfrac{1}{2}}  \right) \frac{R_E^4}{d}$ } \\ 
            \rule{0pt}{3ex} 
             $2g+h$        & \scalebox{.75}{$\left(\dfrac{-5.419\times 10^{-5}}{(\frac{a}{R_E})^{4}}-(3.448\times10^{-9})\frac{a}{R_E} \right) \frac{1}{R_E^2}$ }  &\scalebox{.75}{ $\left( \dfrac{-1.281\times 10^{-6}}{(\frac{a}{R_E})^{\sfrac{9}{2}}}-(1.551\times10^{-11})(\frac{a}{R_E})^{\sfrac{1}{2}}  \right) \frac{R_E^4}{d}$} \\ 
            \rule{0pt}{3ex} 
            $2g$        &\scalebox{.75}{ $\left(\dfrac{-4.060\times 10^{-4}}{(\frac{a}{R_E})^{4}}-(2.329\times10^{-8})\frac{a}{R_E} \right) \frac{1}{R_E^2}$ }  & \scalebox{.75}{$\left( \dfrac{-1.042\times 10^{-5}}{(\frac{a}{R_E})^{\sfrac{9}{2}}}-(1.087\times10^{-10})(\frac{a}{R_E})^{\sfrac{1}{2}}  \right) \frac{R_E^4}{d}$} \\
            \rule{0pt}{3ex} 
            $2g-h$        &\scalebox{.75}{ $\left(\dfrac{-7.984\times 10^{-4}}{(\frac{a}{R_E})^{4}}-(4.080\times10^{-8})\frac{a}{R_E} \right) \frac{1}{R_E^2}$}   & \scalebox{.75}{$\left( \dfrac{-2.213\times 10^{-5}}{(\frac{a}{R_E})^{\sfrac{9}{2}}}-(3.923\times10^{-10})(\frac{a}{R_E})^{\sfrac{1}{2}}  \right) \frac{R_E^4}{d}$ }\\
            \rule{0pt}{3ex} 
            $g-h$        & \scalebox{.75}{ $\left(\dfrac{-1.282\times 10^{-3}}{(\frac{a}{R_E})^{4}}-(5.856\times10^{-8})\frac{a}{R_E} \right) \frac{1}{R_E^2}$  } & \scalebox{.75}{$\left( \dfrac{-3.783\times 10^{-5}}{(\frac{a}{R_E})^{\sfrac{9}{2}}}-(8.361\times10^{-10})(\frac{a}{R_E})^{\sfrac{1}{2}}  \right) \frac{R_E^4}{d}$} \\
            \Xhline{1.15pt}  
            \end{tabular}}
        \end{adjustbox}
        \end{table}
            
            \begin{table}
            \centering
            \caption{Coefficients of $J_F^2 X^2$,  $J_F^2 Y^2$.}
            \begin{adjustbox}{width=\textwidth}
            {\renewcommand{\arraystretch}{2.5}
            \begin{tabular}{|c|c|c|}
            \Xhline{1.15pt}  
            Resonance   & Coeff. of $J_F^2 X^2$ &  Coeff. of $J_F^2 X^2$\\
            \Xhline{1.15pt}  
            $g+h$        & \scalebox{.75}{$\left(\dfrac{-1.895\times 10^{-5}}{(\frac{a}{R_E})^{\sfrac{9}{2}}}-(5.155\times10^{-10})(\frac{a}{R_E})^{\sfrac{1}{2}} \right) \frac{R_E^4}{d} $ }  &
             \scalebox{.75}{ $\left(\dfrac{-1.895\times 10^{-5}}{(\frac{a}{R_E})^{\sfrac{9}{2}}}-(5.718\times10^{-10})(\frac{a}{R_E})^{\sfrac{1}{2}}\right) \frac{R_E^4}{d} $} \\ 
            \rule{0pt}{3ex} 
             $2g+h$        & \scalebox{.75}{$\left(\dfrac{-1.369\times 10^{-5}}{(\frac{a}{R_E})^{\sfrac{9}{2}}}-(1.074\times10^{-11})(\frac{a}{R_E})^{\sfrac{1}{2}} \right) \frac{R_E^4}{d}$ }  & \scalebox{.75}{ $\left(\dfrac{-1.895\times 10^{-5}}{(\frac{a}{R_E})^{\sfrac{9}{2}}}-(1.074\times10^{-9})(\frac{a}{R_E})^{\sfrac{1}{2}} \right) \frac{R_E^4}{d}$ }   \\ 
            \rule{0pt}{3ex} 
             $2g$        & \scalebox{.75}{$\left(\dfrac{-1.895\times 10^{-5}}{(\frac{a}{R_E})^{\sfrac{9}{2}}}-(1.087\times10^{-9})(\frac{a}{R_E})^{\sfrac{1}{2}} \right) \frac{R_E^4}{d}$ }  &
            \scalebox{.75}{ $\left(\dfrac{-1.895\times 10^{-5}}{(\frac{a}{R_E})^{\sfrac{9}{2}}}\right) \frac{R_E^4}{d}$}\\
            \rule{0pt}{3ex} 
             $2g-h$        & \scalebox{.75}{$\left(\dfrac{-1.895\times 10^{-5}}{(\frac{a}{R_E})^{\sfrac{9}{2}}}-(5.464\times10^{-10})(\frac{a}{R_E})^{\sfrac{1}{2}} \right) \frac{R_E^4}{d}$ }  &
             \scalebox{.75}{ $\left(\dfrac{-1.895\times 10^{-5}}{(\frac{a}{R_E})^{\sfrac{9}{2}}}-(5.409\times10^{-10})(\frac{a}{R_E})^{\sfrac{1}{2}}\right) \frac{R_E^4}{d} $}\\
            \rule{0pt}{3ex} 
              $g-h$        &\scalebox{.75}{ $\left(\dfrac{-1.895\times 10^{-5}}{(\frac{a}{R_E})^{\sfrac{9}{2}}}-(5.155\times10^{-10})(\frac{a}{R_E})^{\sfrac{1}{2}} \right) \frac{R_E^4}{d}$}   &
            \scalebox{.75}{$\left( \dfrac{-1.895\times 10^{-5}}{(\frac{a}{R_E})^{\sfrac{9}{2}}}-(5.718\times10^{-10})(\frac{a}{R_E})^{\sfrac{1}{2}} \right) \frac{R_E^4}{d}$} \\
            \Xhline{1.15pt}  
            \end{tabular}}
        \end{adjustbox}
        \end{table}
            
        \begin{table}
            \centering
            \caption{Coefficients of $J_F^2 X^4$,  $J_F^2 Y^4$.}
        \begin{adjustbox}{width=\textwidth}
                {\renewcommand{\arraystretch}{2.5}
            \begin{tabular}{|c|c|c|}
            \Xhline{1.15pt}  
            Resonance   & Coeff. of $J_F^2 X^4$ &  Coeff. of $J_F^2 X^4$\\
            \Xhline{1.15pt}  
            $g+h$        &\scalebox{.75}{  $\left(\dfrac{2.028\times 10^{-5}}{(\frac{a}{R_E})^{\sfrac{9}{2}}}+(3.863\times10^{-12})(\frac{a}{R_E})^{\sfrac{1}{2}}  \right) \frac{R_E^4}{d}$}   &
             \scalebox{.75}{  $\left(\dfrac{2.028\times 10^{-5}}{(\frac{a}{R_E})^{\sfrac{9}{2}}}+(3.403\times10^{-11})(\frac{a}{R_E})^{\sfrac{1}{2}}  \right) \frac{R_E^4}{d}$ }\\ 
            \rule{0pt}{3ex} 
              $2g+h$  &\scalebox{.75}{  $\left(\dfrac{2.227\times 10^{-5}}{(\frac{a}{R_E})^{\sfrac{9}{2}}}+(1.383\times10^{-10})(\frac{a}{R_E})^{\sfrac{1}{2}}  \right) \frac{R_E^4}{d}$}   &
             \scalebox{.75}{ $\left(\dfrac{2.227\times 10^{-5}}{(\frac{a}{R_E})^{\sfrac{9}{2}}}+(2.285\times10^{-10})(\frac{a}{R_E})^{\sfrac{1}{2}} \right) \frac{R_E^4}{d}$  } \\ 
            \rule{0pt}{3ex} 
            $2g$        & \scalebox{.75}{ $\left(\dfrac{2.543\times 10^{-5}}{(\frac{a}{R_E})^{\sfrac{9}{2}}}+(7.293\times10^{-10})(\frac{a}{R_E})^{\sfrac{1}{2}} \right) \frac{R_E^4}{d}$}   &
             \scalebox{.75}{ $\left(\dfrac{2.543\times 10^{-5}}{(\frac{a}{R_E})^{\sfrac{9}{2}}} \right) \frac{R_E^4}{d}$}\\
            \rule{0pt}{3ex} 
            $2g-h$        &\scalebox{.75}{  $\left(\dfrac{2.543\times 10^{-5}}{(\frac{a}{R_E})^{\sfrac{9}{2}}}+(7.293\times10^{-10})(\frac{a}{R_E})^{\sfrac{1}{2}} \right) \frac{R_E^4}{d}$ }  &
             \scalebox{.75}{ $\left( \dfrac{2.543\times 10^{-5}}{(\frac{a}{R_E})^{\sfrac{9}{2}}} \right) \frac{R_E^4}{d}$}\\
            \rule{0pt}{3ex} 
              $g-h$        & \scalebox{.75}{ $\left(\dfrac{3.544\times 10^{-5}}{(\frac{a}{R_E})^{\sfrac{9}{2}} }+(3.544\times10^{-10})(\frac{a}{R_E})^{\sfrac{1}{2}}\right) \frac{R_E^4}{d}$  } &
             \scalebox{.75}{ $\left(\dfrac{3.544\times 10^{-5}}{(\frac{a}{R_E})^{\sfrac{9}{2}}}+(8.064\times10^{-10})(\frac{a}{R_E})^{\sfrac{1}{2}} \right) \frac{R_E^4}{d}$ }\\
            \Xhline{1.15pt}  
            \end{tabular}}
        \end{adjustbox}
        \end{table}

\end{appendices}
\clearpage

\bibliography{refs}


\begin{thebibliography}{31}
\ifx \bisbn   \undefined \def \bisbn  #1{ISBN #1}\fi
\ifx \binits  \undefined \def \binits#1{#1}\fi
\ifx \bauthor  \undefined \def \bauthor#1{#1}\fi
\ifx \batitle  \undefined \def \batitle#1{#1}\fi
\ifx \bjtitle  \undefined \def \bjtitle#1{#1}\fi
\ifx \bvolume  \undefined \def \bvolume#1{\textbf{#1}}\fi
\ifx \byear  \undefined \def \byear#1{#1}\fi
\ifx \bissue  \undefined \def \bissue#1{#1}\fi
\ifx \bfpage  \undefined \def \bfpage#1{#1}\fi
\ifx \blpage  \undefined \def \blpage #1{#1}\fi
\ifx \burl  \undefined \def \burl#1{\textsf{#1}}\fi
\ifx \doiurl  \undefined \def \doiurl#1{\url{https://doi.org/#1}}\fi
\ifx \betal  \undefined \def \betal{\textit{et al.}}\fi
\ifx \binstitute  \undefined \def \binstitute#1{#1}\fi
\ifx \binstitutionaled  \undefined \def \binstitutionaled#1{#1}\fi
\ifx \bctitle  \undefined \def \bctitle#1{#1}\fi
\ifx \beditor  \undefined \def \beditor#1{#1}\fi
\ifx \bpublisher  \undefined \def \bpublisher#1{#1}\fi
\ifx \bbtitle  \undefined \def \bbtitle#1{#1}\fi
\ifx \bedition  \undefined \def \bedition#1{#1}\fi
\ifx \bseriesno  \undefined \def \bseriesno#1{#1}\fi
\ifx \blocation  \undefined \def \blocation#1{#1}\fi
\ifx \bsertitle  \undefined \def \bsertitle#1{#1}\fi
\ifx \bsnm \undefined \def \bsnm#1{#1}\fi
\ifx \bsuffix \undefined \def \bsuffix#1{#1}\fi
\ifx \bparticle \undefined \def \bparticle#1{#1}\fi
\ifx \barticle \undefined \def \barticle#1{#1}\fi
\bibcommenthead
\ifx \bconfdate \undefined \def \bconfdate #1{#1}\fi
\ifx \botherref \undefined \def \botherref #1{#1}\fi
\ifx \url \undefined \def \url#1{\textsf{#1}}\fi
\ifx \bchapter \undefined \def \bchapter#1{#1}\fi
\ifx \bbook \undefined \def \bbook#1{#1}\fi
\ifx \bcomment \undefined \def \bcomment#1{#1}\fi
\ifx \oauthor \undefined \def \oauthor#1{#1}\fi
\ifx \citeauthoryear \undefined \def \citeauthoryear#1{#1}\fi
\ifx \endbibitem  \undefined \def \endbibitem {}\fi
\ifx \bconflocation  \undefined \def \bconflocation#1{#1}\fi
\ifx \arxivurl  \undefined \def \arxivurl#1{\textsf{#1}}\fi
\csname PreBibitemsHook\endcsname

\bibitem{daquin2021deep}
\begin{barticle}
\bauthor{\bsnm{Daquin}, \binits{J.}},
\bauthor{\bsnm{Legnaro}, \binits{E.}},
\bauthor{\bsnm{Gkolias}, \binits{I.}},
\bauthor{\bsnm{Efthymiopoulos}, \binits{C.}}:
\batitle{A deep dive into the $2g+h$ resonance: separatrices, manifolds and
  phase space structure of navigation satellites}.
\bjtitle{Celestial Mechanics and Dynamical Astronomy}
\bvolume{134}(\bissue{1}),
\bfpage{1}--\blpage{31}
(\byear{2022})
\end{barticle}
\endbibitem

\bibitem{hughes1980earth}
\begin{barticle}
\bauthor{\bsnm{Hughes}, \binits{S.}}:
\batitle{Earth satellite orbits with resonant lunisolar perturbations i.
  resonances dependent only on inclination}.
\bjtitle{Proceedings of the Royal Society of London. A. Mathematical and
  Physical Sciences}
\bvolume{372}(\bissue{1749}),
\bfpage{243}--\blpage{264}
(\byear{1980})
\end{barticle}
\endbibitem

\bibitem{rossi2008resonant}
\begin{barticle}
\bauthor{\bsnm{Rossi}, \binits{A.}}:
\batitle{Resonant dynamics of medium earth orbits: space debris issues}.
\bjtitle{Celestial Mechanics and Dynamical Astronomy}
\bvolume{100}(\bissue{4}),
\bfpage{267}--\blpage{286}
(\byear{2008})
\end{barticle}
\endbibitem

\bibitem{rosengren2015chaos}
\begin{barticle}
\bauthor{\bsnm{Rosengren}, \binits{A.J.}},
\bauthor{\bsnm{Alessi}, \binits{E.M.}},
\bauthor{\bsnm{Rossi}, \binits{A.}},
\bauthor{\bsnm{Valsecchi}, \binits{G.B.}}:
\batitle{Chaos in navigation satellite orbits caused by the perturbed motion of
  the moon}.
\bjtitle{Monthly Notices of the Royal Astronomical Society}
\bvolume{449}(\bissue{4}),
\bfpage{3522}--\blpage{3526}
(\byear{2015})
\end{barticle}
\endbibitem

\bibitem{daquin2016dynamical}
\begin{barticle}
\bauthor{\bsnm{Daquin}, \binits{J.}},
\bauthor{\bsnm{Rosengren}, \binits{A.J.}},
\bauthor{\bsnm{Alessi}, \binits{E.M.}},
\bauthor{\bsnm{Deleflie}, \binits{F.}},
\bauthor{\bsnm{Valsecchi}, \binits{G.B.}},
\bauthor{\bsnm{Rossi}, \binits{A.}}:
\batitle{The dynamical structure of the meo region: long-term stability, chaos,
  and transport}.
\bjtitle{Celestial Mechanics and Dynamical Astronomy}
\bvolume{124}(\bissue{4}),
\bfpage{335}--\blpage{366}
(\byear{2016})
\end{barticle}
\endbibitem

\bibitem{celletti2016study}
\begin{barticle}
\bauthor{\bsnm{Celletti}, \binits{A.}},
\bauthor{\bsnm{Gale{\c{s}}}, \binits{C.B.}}:
\batitle{A study of the lunisolar secular resonance $2\dot{\omega} +
  \dot{\Omega}= 0$}.
\bjtitle{Frontiers in Astronomy and Space Sciences}
\bvolume{3},
\bfpage{11}
(\byear{2016})
\end{barticle}
\endbibitem

\bibitem{gkolias2016order}
\begin{barticle}
\bauthor{\bsnm{Gkolias}, \binits{I.}},
\bauthor{\bsnm{Daquin}, \binits{J.}},
\bauthor{\bsnm{Gachet}, \binits{F.}},
\bauthor{\bsnm{Rosengren}, \binits{A.J.}}:
\batitle{From order to chaos in earth satellite orbits}.
\bjtitle{The Astronomical Journal}
\bvolume{152}(\bissue{5}),
\bfpage{119}
(\byear{2016})
\end{barticle}
\endbibitem

\bibitem{celletti2017analytical}
\begin{barticle}
\bauthor{\bsnm{Celletti}, \binits{A.}},
\bauthor{\bsnm{Gale{\c{s}}}, \binits{C.}},
\bauthor{\bsnm{Pucacco}, \binits{G.}},
\bauthor{\bsnm{Rosengren}, \binits{A.J.}}:
\batitle{Analytical development of the lunisolar disturbing function and the
  critical inclination secular resonance}.
\bjtitle{Celestial Mechanics and Dynamical Astronomy}
\bvolume{127}(\bissue{3}),
\bfpage{259}--\blpage{283}
(\byear{2017})
\end{barticle}
\endbibitem

\bibitem{celletti2020resonances}
\begin{barticle}
\bauthor{\bsnm{Celletti}, \binits{A.}},
\bauthor{\bsnm{Gales}, \binits{C.}},
\bauthor{\bsnm{Lhotka}, \binits{C.}}:
\batitle{Resonances in the earth’s space environment}.
\bjtitle{Communications in Nonlinear Science and Numerical Simulation}
\bvolume{84},
\bfpage{105185}
(\byear{2020})
\end{barticle}
\endbibitem

\bibitem{alessi2021dynamical}
\begin{barticle}
\bauthor{\bsnm{Alessi}, \binits{E.M.}},
\bauthor{\bsnm{Buzzoni}, \binits{A.}},
\bauthor{\bsnm{Daquin}, \binits{J.}},
\bauthor{\bsnm{Carbognani}, \binits{A.}},
\bauthor{\bsnm{Tommei}, \binits{G.}}:
\batitle{Dynamical properties of the molniya satellite constellation: Long-term
  evolution of orbital eccentricity}.
\bjtitle{Acta Astronautica}
\bvolume{179},
\bfpage{659}--\blpage{669}
(\byear{2021})
\end{barticle}
\endbibitem

\bibitem{daquin2021dynamical}
\begin{barticle}
\bauthor{\bsnm{Daquin}, \binits{J.}},
\bauthor{\bsnm{Alessi}, \binits{E.M.}},
\bauthor{\bsnm{O’Leary}, \binits{J.}},
\bauthor{\bsnm{Lemaitre}, \binits{A.}},
\bauthor{\bsnm{Buzzoni}, \binits{A.}}:
\batitle{Dynamical properties of the molniya satellite constellation: long-term
  evolution of the semi-major axis}.
\bjtitle{Nonlinear Dynamics}
\bvolume{105}(\bissue{3}),
\bfpage{2081}--\blpage{2103}
(\byear{2021})
\end{barticle}
\endbibitem

\bibitem{talu2021dominant}
\begin{barticle}
\bauthor{\bsnm{Talu}, \binits{T.}},
\bauthor{\bsnm{Alessi}, \binits{E.M.}},
\bauthor{\bsnm{Tommei}, \binits{G.}}:
\batitle{On the dominant lunisolar perturbations for long-term eccentricity
  variation: The case of molniya satellite orbits}.
\bjtitle{Universe}
\bvolume{7}(\bissue{12}),
\bfpage{482}
(\byear{2021})
\end{barticle}
\endbibitem

\bibitem{breiter2001lunisolar}
\begin{barticle}
\bauthor{\bsnm{Breiter}, \binits{S.}}:
\batitle{Lunisolar resonances revisited}.
\bjtitle{Celestial Mechanics and Dynamical Astronomy}
\bvolume{81}(\bissue{1-2}),
\bfpage{81}--\blpage{91}
(\byear{2001})
\end{barticle}
\endbibitem

\bibitem{chao2004long}
\begin{barticle}
\bauthor{\bsnm{Chao}, \binits{C.}},
\bauthor{\bsnm{Gick}, \binits{R.}}:
\batitle{Long-term evolution of navigation satellite orbits:
  Gps/glonass/galileo}.
\bjtitle{Advances in Space Research}
\bvolume{34}(\bissue{5}),
\bfpage{1221}--\blpage{1226}
(\byear{2004})
\end{barticle}
\endbibitem

\bibitem{alessi2014effectiveness}
\begin{barticle}
\bauthor{\bsnm{Alessi}, \binits{E.M.}},
\bauthor{\bsnm{Rossi}, \binits{A.}},
\bauthor{\bsnm{Valsecchi}, \binits{G.}},
\bauthor{\bsnm{Anselmo}, \binits{L.}},
\bauthor{\bsnm{Pardini}, \binits{C.}},
\bauthor{\bsnm{Colombo}, \binits{C.}},
\bauthor{\bsnm{Lewis}, \binits{H.}},
\bauthor{\bsnm{Daquin}, \binits{J.}},
\bauthor{\bsnm{Deleflie}, \binits{F.}},
\bauthor{\bsnm{Vasile}, \binits{M.}}, \betal:
\batitle{Effectiveness of gnss disposal strategies}.
\bjtitle{Acta Astronautica}
\bvolume{99},
\bfpage{292}--\blpage{302}
(\byear{2014})
\end{barticle}
\endbibitem

\bibitem{alessi2016numerical}
\begin{barticle}
\bauthor{\bsnm{Alessi}, \binits{E.}},
\bauthor{\bsnm{Deleflie}, \binits{F.}},
\bauthor{\bsnm{Rosengren}, \binits{A.}},
\bauthor{\bsnm{Rossi}, \binits{A.}},
\bauthor{\bsnm{Valsecchi}, \binits{G.}},
\bauthor{\bsnm{Daquin}, \binits{J.}},
\bauthor{\bsnm{Merz}, \binits{K.}}:
\batitle{A numerical investigation on the eccentricity growth of gnss disposal
  orbits}.
\bjtitle{Celestial Mechanics and Dynamical Astronomy}
\bvolume{125}(\bissue{1}),
\bfpage{71}--\blpage{90}
(\byear{2016})
\end{barticle}
\endbibitem

\bibitem{armellin2018optimal}
\begin{barticle}
\bauthor{\bsnm{Armellin}, \binits{R.}},
\bauthor{\bsnm{San-Juan}, \binits{J.F.}}:
\batitle{Optimal earth’s reentry disposal of the galileo constellation}.
\bjtitle{Advances in Space Research}
\bvolume{61}(\bissue{4}),
\bfpage{1097}--\blpage{1120}
(\byear{2018})
\end{barticle}
\endbibitem

\bibitem{skoulidou2019medium}
\begin{barticle}
\bauthor{\bsnm{Skoulidou}, \binits{D.K.}},
\bauthor{\bsnm{Rosengren}, \binits{A.J.}},
\bauthor{\bsnm{Tsiganis}, \binits{K.}},
\bauthor{\bsnm{Voyatzis}, \binits{G.}}:
\batitle{Medium earth orbit dynamical survey and its use in passive debris
  removal}.
\bjtitle{Advances in Space Research}
\bvolume{63}(\bissue{11}),
\bfpage{3646}--\blpage{3674}
(\byear{2019})
\end{barticle}
\endbibitem

\bibitem{froeschle1997fast}
\begin{barticle}
\bauthor{\bsnm{Froeschl{\'e}}, \binits{C.}},
\bauthor{\bsnm{Lega}, \binits{E.}},
\bauthor{\bsnm{Gonczi}, \binits{R.}}:
\batitle{Fast lyapunov indicators. application to asteroidal motion}.
\bjtitle{Celestial Mechanics and Dynamical Astronomy}
\bvolume{67}(\bissue{1}),
\bfpage{41}--\blpage{62}
(\byear{1997})
\end{barticle}
\endbibitem

\bibitem{guzzo2002numerical}
\begin{barticle}
\bauthor{\bsnm{Guzzo}, \binits{M.}},
\bauthor{\bsnm{Lega}, \binits{E.}},
\bauthor{\bsnm{Froeschl{\'e}}, \binits{C.}}:
\batitle{On the numerical detection of the effective stability of chaotic
  motions in quasi-integrable systems}.
\bjtitle{Physica D: Nonlinear Phenomena}
\bvolume{163}(\bissue{1-2}),
\bfpage{1}--\blpage{25}
(\byear{2002})
\end{barticle}
\endbibitem

\bibitem{kozai1962secular}
\begin{barticle}
\bauthor{\bsnm{Kozai}, \binits{Y.}}:
\batitle{Secular perturbations of asteroids with high inclination and
  eccentricity}.
\bjtitle{The Astronomical Journal}
\bvolume{67},
\bfpage{591}--\blpage{598}
(\byear{1962})
\end{barticle}
\endbibitem

\bibitem{lidov1961evolution}
\begin{barticle}
\bauthor{\bsnm{Lidov}, \binits{M.}}:
\batitle{Evolution of artificial planetary satellites under the action of
  gravitational perturbations due to external bodies}.
\bjtitle{Iskusstviennye Sputniki Zemli}
\bvolume{8},
\bfpage{5}--\blpage{45}
(\byear{1961})
\end{barticle}
\endbibitem

\bibitem{kaula1962development}
\begin{barticle}
\bauthor{\bsnm{Kaula}, \binits{W.M.}}:
\batitle{Development of the lunar and solar disturbing functions for a close
  satellite}.
\bjtitle{Astronomical Journal}
\bvolume{67}(\bissue{5}),
\bfpage{300}
(\byear{1962}).
Accessed 2022-08-05
\end{barticle}
\endbibitem

\bibitem{celletti2016bifurcation}
\begin{barticle}
\bauthor{\bsnm{Celletti}, \binits{A.}},
\bauthor{\bsnm{Gales}, \binits{C.}},
\bauthor{\bsnm{Pucacco}, \binits{G.}}:
\batitle{Bifurcation of lunisolar secular resonances for space debris orbits}.
\bjtitle{SIAM Journal on Applied Dynamical Systems}
\bvolume{15}(\bissue{3}),
\bfpage{1352}--\blpage{1383}
(\byear{2016})
\end{barticle}
\endbibitem

\bibitem{gkolias2019chaotic}
\begin{barticle}
\bauthor{\bsnm{Gkolias}, \binits{I.}},
\bauthor{\bsnm{Daquin}, \binits{J.}},
\bauthor{\bsnm{Skoulidou}, \binits{D.K.}},
\bauthor{\bsnm{Tsiganis}, \binits{K.}},
\bauthor{\bsnm{Efthymiopoulos}, \binits{C.}}:
\batitle{Chaotic transport of navigation satellites}.
\bjtitle{Chaos: An Interdisciplinary Journal of Nonlinear Science}
\bvolume{29}(\bissue{10}),
\bfpage{101106}
(\byear{2019})
\end{barticle}
\endbibitem

\bibitem{henrard1983second}
\begin{barticle}
\bauthor{\bsnm{Henrard}, \binits{J.}},
\bauthor{\bsnm{Lemaitre}, \binits{A.}}:
\batitle{A second fundamental model for resonance}.
\bjtitle{Celestial mechanics}
\bvolume{30}(\bissue{2}),
\bfpage{197}--\blpage{218}
(\byear{1983})
\end{barticle}
\endbibitem

\bibitem{lei2021secular}
\begin{botherref}
\oauthor{\bsnm{Lei}, \binits{H.}},
\oauthor{\bsnm{Ortore}, \binits{E.}},
\oauthor{\bsnm{Circi}, \binits{C.}}:
Secular dynamics of navigation satellites in the meo and gso regions.
Astrodynamics,
1--18
(2021)
\end{botherref}
\endbibitem

\bibitem{rosengren2016galileo}
\begin{barticle}
\bauthor{\bsnm{Rosengren}, \binits{A.J.}},
\bauthor{\bsnm{Daquin}, \binits{J.}},
\bauthor{\bsnm{Tsiganis}, \binits{K.}},
\bauthor{\bsnm{Alessi}, \binits{E.M.}},
\bauthor{\bsnm{Deleflie}, \binits{F.}},
\bauthor{\bsnm{Rossi}, \binits{A.}},
\bauthor{\bsnm{Valsecchi}, \binits{G.B.}}:
\batitle{Galileo disposal strategy: stability, chaos and predictability}.
\bjtitle{Monthly Notices of the Royal Astronomical Society}
\bvolume{464}(\bissue{4}),
\bfpage{4063}--\blpage{4076}
(\byear{2016})
\end{barticle}
\endbibitem

\bibitem{allan1964long}
\begin{barticle}
\bauthor{\bsnm{Allan}, \binits{R.}},
\bauthor{\bsnm{Cook}, \binits{G.}}:
\batitle{The long-period motion of the plane of a distant circular orbit}.
\bjtitle{Proceedings of the Royal Society of London. Series A. Mathematical and
  Physical Sciences}
\bvolume{280}(\bissue{1380}),
\bfpage{97}--\blpage{109}
(\byear{1964})
\end{barticle}
\endbibitem

\bibitem{kudielka1997equilibria}
\begin{bchapter}
\bauthor{\bsnm{Kudielka}, \binits{V.W.}}:
\bctitle{Equilibria bifurcations of satellite orbits}.
In: \beditor{\bsnm{Dvorak}, \binits{R.}},
\beditor{\bsnm{Henrard}, \binits{J.}} (eds.)
\bbtitle{The Dynamical Behaviour of Our Planetary System},
pp. \bfpage{243}--\blpage{255}.
\bpublisher{Springer},
\blocation{Dordrecht}
(\byear{1997})
\end{bchapter}
\endbibitem

\bibitem{tremaine2009satellite}
\begin{barticle}
\bauthor{\bsnm{Tremaine}, \binits{S.}},
\bauthor{\bsnm{Touma}, \binits{J.}},
\bauthor{\bsnm{Namouni}, \binits{F.}}:
\batitle{Satellite dynamics on the laplace surface}.
\bjtitle{The astronomical journal}
\bvolume{137}(\bissue{3}),
\bfpage{3706}
(\byear{2009})
\end{barticle}
\endbibitem

\end{thebibliography}


\end{document}